\documentclass{aa}  

\usepackage{graphicx}
\usepackage{txfonts}
\usepackage{silence}
\usepackage{subcaption} %
\usepackage{amsmath,amssymb}
\usepackage{hyperref}
\usepackage{booktabs}
\usepackage[dvipsnames]{xcolor}
\definecolor{myblue}{HTML}{1F77B4}
\definecolor{mygreen}{HTML}{2CA02C}
\definecolor{myred}{HTML}{D62728}
\definecolor{mymagenta}{HTML}{D33682}
\definecolor{codepurple}{HTML}{C42043}

\hypersetup{
    unicode=true,           %
    colorlinks=true,        %
    linkcolor=myred,        %
    citecolor=myblue,       %
    filecolor=magenta,      %
    urlcolor=mygreen        %
}

\usepackage{bookmark}%

\usepackage{graphicx}	%
\usepackage{amsmath}	%

\usepackage{longtable}
\usepackage{booktabs}
\usepackage{xcolor}
\usepackage{xspace}

\usepackage[switch]{lineno}

\newcommand{\xdos}{SN~2021adxl\xspace}
\newcommand{\mdot}{$\dot{M}$\xspace}

\newcommand{\xdosZTF}{ZTF21ackxdos\xspace}
\newcommand{\xshooter}{X-shooter\xspace}

\newcommand{\halpha}{H$\alpha$\xspace}
\newcommand{\lalpha}{Ly$\alpha$\xspace}
\newcommand{\hbeta}{H$\beta$\xspace}
\newcommand{\hgamma}{H$\gamma$\xspace}

\newcommand{\kms}{~km s$^{-1}$\xspace}
\newcommand{\msun}{M$_{\odot}$\xspace}
\newcommand{\msunperyr}{\xspace M$_{\odot}$ yr$^{-1}$\xspace}

\newcommand{\magperhundredday}{mag / 100 d\xspace}
\newcommand{\software}[1]{\textsc{#1}}

\begin{document}

\title{\xdos: A luminous nearby interacting supernova in an extremely low metallicity environment}

\titlerunning{The nearby luminous interacting \xdos}

\author{
S. J. Brennan\inst{1} \href{https://orcid.org/0000-0003-1325-6235}{\includegraphics[scale=0.05]{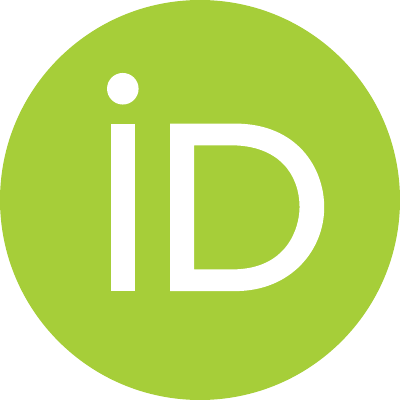}}\and
S. Schulze\inst{2,3} \href{https://orcid.org/0000-0001-6797-1889}{\includegraphics[scale=0.05]{figures/ORCIDiD.pdf}}\and
R. Lunnan\inst{1} \href{https://orcid.org/0000-0001-9454-4639}{\includegraphics[scale=0.05]{figures/ORCIDiD.pdf}}\and
J. Sollerman\inst{1} \href{https://orcid.org/0000-0003-1546-6615}{\includegraphics[scale=0.05]{figures/ORCIDiD.pdf}}\and
L. Yan\inst{4} \href{https://orcid.org/0000-0003-1710-9339}{\includegraphics[scale=0.05]{figures/ORCIDiD.pdf}}\and
C. Fransson\inst{1} \href{https://orcid.org/0000-0001-8532-3594}{\includegraphics[scale=0.05]{figures/ORCIDiD.pdf}}\and
I. Irani\inst{5} \href{https://orcid.org/0000-0002-7996-8780}{\includegraphics[scale=0.05]{figures/ORCIDiD.pdf}}\and
J. Melinder\inst{1} \href{https://orcid.org/0000-0003-0470-8754}{\includegraphics[scale=0.05]{figures/ORCIDiD.pdf}}\and
T.-W. Chen\inst{6} \href{https://orcid.org/0000-0002-1066-6098}{\includegraphics[scale=0.05]{figures/ORCIDiD.pdf}}\and
K. De\inst{7} \href{https://orcid.org/0000-0002-8989-0542}{\includegraphics[scale=0.05]{figures/ORCIDiD.pdf}}\and
C. Fremling\inst{4} \href{https://orcid.org/0000-0002-4223-103X}{\includegraphics[scale=0.05]{figures/ORCIDiD.pdf}}\and
Y.-L. Kim\inst{8} \href{https://orcid.org/0000-0002-1031-0796}{\includegraphics[scale=0.05]{figures/ORCIDiD.pdf}}\and
D. Perley\inst{9} \href{https://orcid.org/0000-0001-8472-1996}{\includegraphics[scale=0.05]{figures/ORCIDiD.pdf}}\and
P. J. Pessi\inst{1} \href{https://orcid.org/0000-0001-8532-3594}{\includegraphics[scale=0.05]{figures/ORCIDiD.pdf}}\and
A. J. Drake\inst{10} \href{https://orcid.org}{\includegraphics[scale=0.05]{figures/ORCIDiD.pdf}}\and
M. J. Graham\inst{10} \href{https://orcid.org/0000-0002-3168-0139}{\includegraphics[scale=0.05]{figures/ORCIDiD.pdf}}\and
R. R. Laher\inst{11} \href{https://orcid.org/0000-0003-2451-5482}{\includegraphics[scale=0.05]{figures/ORCIDiD.pdf}}\and
F. J. Masci\inst{11} \href{https://orcid.org/0000-0002-8532-9395}{\includegraphics[scale=0.05]{figures/ORCIDiD.pdf}}\and
J. Purdum\inst{4} \href{https://orcid.org/0000-0003-1227-3738}{\includegraphics[scale=0.05]{figures/ORCIDiD.pdf}}\and
H. Rodriguez\inst{4} \href{https://orcid.org}{\includegraphics[scale=0.05]{figures/ORCIDiD.pdf}}
}

\institute{
The Oskar Klein Centre, Department of Astronomy, Stockholm University, AlbaNova, SE-10691 Stockholm, Sweden 
  \and 
 Center for Interdisciplinary Exploration and Research in Astrophysics (CIERA), Northwestern University, 1800 Sherman Ave., Evanston, IL 60201, USA 
  \and 
 The Oskar Klein Centre, Department of Physics, Stockholm University, Albanova University Center, 106 91 Stockholm, Sweden 
  \and 
 Caltech Optical Observatories, California Institute of Technology, Pasadena, CA 91125, USA 
  \and 
 Department of Particle Physics and Astrophysics, Weizmann Institute of Science, 234 Herzl St, 7610001 Rehovot, Israel 
  \and 
 Graduate Institute of Astronomy, National Central University, 300 Jhongda Road, 32001 Jhongli, Taiwan 
  \and 
 MIT-Kavli Institute for Astrophysics and Space Research, 77 Massachusetts Ave., Cambridge, MA 02139, USA 
  \and 
 Department of Physics, Lancaster University, Lancaster LA1 4YW, UK 
  \and 
 Astrophysics Research Institute, Liverpool John Moores University, IC2, Liverpool Science Park, 146 Brownlow Hill, Liverpool L3 5RF, UK 
  \and 
 Division of Physics, Mathematics and Astronomy, California Institute of Technology, Pasadena, CA 91125, USA 
  \and 
 IPAC, California Institute of Technology, 1200 E. California Blvd, Pasadena, CA 91125, USA 
 }

\abstract
{

\xdos is a slowly evolving, luminous, Type IIn supernova  with asymmetric emission line profiles, similar to the well-studied SN 2010jl. We present extensive optical, near-ultraviolet, and near-infrared photometry and spectroscopy covering $\sim$1.5~years post discovery. \xdos occurred in an unusual environment, atop a vigorously star-forming region which is offset from its host galaxy core. The appearance of \lalpha, \ion{O}{II} , as well as the compact core, would classify the host of \xdos as a Blueberry galaxy, analogous to higher redshift, low metallicity, star-forming dwarf ``Green Pea'' galaxies . Using several abundance indicators, we find a metallicity of the explosion environment of only $\sim0.1~{\rm Z_{\odot}}$, the lowest reported metallicity for a Type IIn SN environment. \xdos reaches a peak magnitude of M${_r}~\approx-20.2$~mag and since discovery, \xdos has faded by only $\sim$4 magnitudes in the $r$ band with a cumulative radiated energy of $\sim1.5\times10^{50}$ erg over 18 months. \xdos shows strong signs of interaction with a complex circumstellar medium, seen by the detection of X-rays, revealed by the detection of coronal emission lines, and through multi-component hydrogen and helium profiles. In order to further understand this interaction, we model the \halpha profile using a Monte-Carlo electron scattering code. The blueshifted high-velocity component is consistent with emission from a radially thin, spherical shell resulting in the broad emission components due to electron scattering. Using the velocity evolution of this emitting shell, we find that the SN ejecta collide with circumstellar material  of at least $\sim$5~\msun  assuming a steady-state mass-loss rate of $\sim4-6\times10^{-3}$ \msunperyr for the first $\sim$200 days of evolution. \xdos was last observed to be slowly declining at $\sim0.01~{\rm mag~ d^{-1}}$, and if this trend continues, \xdos will remain observable after its current solar conjunction. Continuing the observations of \xdos may reveal signatures of dust formation or an infrared excess, similar to that seen for SN~2010jl.
}


\maketitle
\section{Introduction}\label{sec:introduction}

Massive stars ($>8$~\msun) are expected to end their lives as core-collapse supernovae (CCSNe) \citep{Woosley1995,Heger2003,Janka2012,Crowther2012}. All-sky photometric surveys \citep{Bellm2014,Chambers2016,Tonry2018} are finding numerous supernovae (SNe) each night,  but the exact mechanisms by which a massive star spends its final moments is currently unclear \citep[see e.g.,][]{Burrows2021}. Supernovae that display narrow Hydrogen emission lines atop a blue continuum are classified as Type IIn SNe \citep{Schlegel1990,Gal-Yam2007,Ransome2021}. The narrow emission features are suggested to arise from the excitation of a slow-moving circum-stellar medium (CSM) near the progenitor which has been photoionised by the ejecta-CSM interaction.

This CSM is expected to have been ejected by the progenitor in the years to decades before core collapse \citep{Vink2008,Ofek2014}. %
While a star will  eject mass into the surrounding interstellar medium (ISM) throughout its life, in the form of stellar winds with mass-loss rates of $10^{-7} - 10^{-4}$ \msunperyr, studies of the progenitors of Type IIn SNe have reported much higher mass-loss rates of $> 10^{-3}$ \msunperyr \citep{Fransson2014,Moriya2014,Fraser2020,Khatami2023}.  This likely reflects either eruptive mass-loss events, or enhanced steady state mass loss by the progenitor in the decades prior to core-collapse.

The physical mechanism behind such an enhanced mass-loss rate is elusive, mainly due to the inability to directly observe these mass-loss events prior to the supernova explosion. However, this brief period of enhanced mass loss likely influences the photometric evolution of the supernova, including duration and luminosity, as well as the spectroscopic appearance of emission/absorption line profiles and their evolution \citep[for example see][ for the case of a disk/torus]{Kurfurst2019,Suzuki2019}. These progenitors must have mass-loss rates not typically observed during the main sequence. It is suggested that the high mass-loss rates are similar to those of massive ($>$ 25~ \msun)  luminous blue variable (LBV) stars  \citep[e.g.][]{Taddia2013,Weis2020} and some Type IIn SNe have been associated with such stars \citep[e.g. SN~2005gl; ][]{Gal-Yam2007}. Alternatively, Type IIn SNe may arise as a result of a superwind phase from a red super-giant (RSG) star \citep{Fransson2002,Smith2009}. Current stellar evolution theory suggests that canonical LBVs are a short transitional phase, between a hydrogen-rich RSG evolving towards a hydrogen-depleted Wolf-Rayet (WR) star \citep{Crowther2007, Langer2012, Ekstrom2012}. The term LBV is typically employed to define a broad phenomenology rather than an evolutionary stage \citep[e.g.][]{Gal-Yam2007,Trundle2008,Dwarkadas2011,Groh2017}. Standard stellar evolution theory predicts that single massive stars which become LBVs do so near the end of, or after, the completion of core hydrogen burning. Thereafter they typically lose their hydrogen-rich envelopes in the LBV phase and become WR stars, where they spend $\sim10^5$~years after which they explode as a CCSN.

To date, there is no clear understanding of how a star may undergo core-collapse at this phase \citep{Chevalier2012,Woosley2017}. However, under certain conditions, a progenitor may have the appearance of a LBV star shortly before core-collapse when the progenitor is rapidly rotating \citep[e.g.][]{Groh2013}. In this scenario, the progenitor likely ejects the majority of its outer envelope during the RSG stage. It should be noted that these stars appear as LBV stars but do not necessarily have the chemical composition of the canonical LBV \citep{Humphreys1994,Weis2020}. Studies of Type IIn SNe show a diverse range of peak magnitudes, ranging from $-16$ to $-23$ magnitudes \citep{Nyholm2020,Ransome2021}. Typically, SNe brighter than $-21$ mag are classified as superluminous supernovae (SLSNe) \citep{Gal-Yam2012,Gal-Yam2019}, although for hydrogen-rich transients it is not clear if there is 
a defined boundary between regular SNe and SLSNe (of Type IIn) or rather an apparent continuity.

\begin{figure}
    \centering
    \includegraphics[width= \columnwidth]{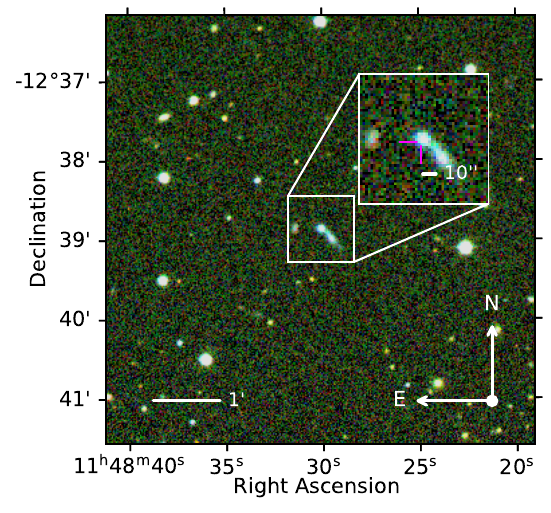}
    \caption{P60/SEDM composite color image ($gri$)  for \xdos (marked in the inset by the magenta crosshair) from the ZTF survey, when the transient was near peak. \xdos is situated atop strong host emission, which can be seen in the inset. }
    \label{fig:finder}
\end{figure}

This paper presents a large dataset for \xdos that enables us to study the late time evolution and environment of a nearby luminous Type IIn SN. A image of \xdos and its environment is given in Fig.~\ref{fig:finder}. \xdos\footnote{Also known as Gaia21fcd, PS22bne, ATLAS23blun, and \xdosZTF.} was discovered by the Zwicky Transient Factory \citep[ZTF; ][]{Graham2019,Bellm2019,Dekany2020} at  R.A. = 11:48:06.940, DEC. = $-$12:38:41.71 (J2000) on 2021-11-08 at a $r$-band magnitude of 14.41 mag \citep{Fremling2021a}.  \xdos was classified by \citealt{De2022} on 2022-02-02 as a Type IIn SN at a redshift of $z=0.018$ (reported wavelengths given in rest frame unless stated otherwise). The host of \xdos , WISEA J114806.88-123841.3, displays a unique morphology known in the literature as a Tadpole Galaxy \citep{Elmegreen2012,Munoz-Tunon2014,Rosado-Belza2019}, discussed further in Sec. \ref{ssec:host}. We assume a Hubble constant $H_{0}=73$\,km\,s$^{-1}$\,Mpc$^{-1}$, $\Omega_{\Lambda}=0.73$ and $\Omega_{\rm M}=0.27$ \citep{Spergel2007}. The corresponding luminosity distance $D_{L}=78.16$~Mpc and a distance modulus of $\mu=34.47\pm0.02$~mag are used for \xdos. We correct for foreground extinction using ${\rm R_V}$=3.1, ${\rm E (B - V)}$ = 0.026  and the extinction law given by \citealt{Cardelli1989}. Unless stated, we do not correct for any potential host galaxy or circumstellar extinction, however we note that the blue colors seen in the spectra of \xdos, as well as the host, do not point towards significant reddening by dust. Additionally the lack of noticeable Na D $\lambda\lambda 5890~5896$ points towards low extinction, although a robust measurement cannot be made to do the strong \ion{He}{I} $\lambda 5876 $ emission. The rising part of the light curve of \xdos was not observed, as the transient was discovered around its peak magnitude after ending solar conjunction. The rest-frame phase is taken with respect to the {\textit r}-band discovery epoch on 2021-11-03 (MJD = 59521). Luminous SNe are intrinsically rare and given the adopted distance of 78~Mpc, \xdos is the nearby Type IIn SN, and just as SN~2010jl \citep[49~Mpc; ][]{Smith2011,Fransson2014,Ofek2014,Jencson2016}, and due to its luminosity and slow evolution, provides a rare opportunity to follow the evolution of such a transient for many years.

This paper is organised as follows: Section \ref{sec:obsdata} provides details of the observational dataset presented in this paper, split into photometric in Sect. \ref{ssec:obsdata_photometry} and spectroscopic data in Sect. \ref{ssec:obsdata_spectra}, as well as gives an overview of the observable properties for \xdos. \xdos occurred in a bright star forming region, and we discuss the host environment of \xdos in Sect. \ref{ssec:host}. A prominent feature of the optical spectra of \xdos is the broad \halpha profile, and we explore the formation of this line in Sect. \ref{sec:halpha_modelling}, as well as any insight into the ejecta-CSM interaction in Sect. \ref{ssec:modelling}. Finally in Sect. \ref{sec:discussion}, using the results from observations, we investigate the metallicity of the host of \xdos and, in general, interacting SNe in Sect. \ref{ssec:metal_poor_host} as well as discuss the possible progenitor of \xdos in Sect. \ref{ssec:progenitor}.

\section{Observations}\label{sec:obsdata}

\subsection{Photometry}\label{ssec:obsdata_photometry}

\begin{figure*}
    \centering
    \includegraphics[width= \textwidth]{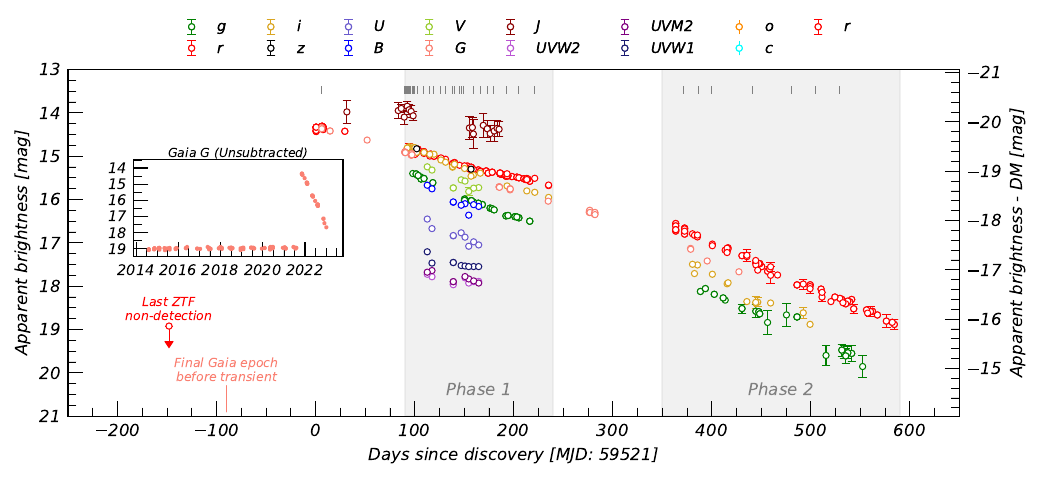}
    \caption{Multi-band lightcurve of \xdos covering $\sim$1.6 years of its post-peak evolution. Highlighted in grey are two distinct phases where \xdos displays two distinct decline slopes, as mentioned in the text. The inset provides Gaia G-band data of \xdos (post-2022), as well as the historic observations of the explosion site (pre-2022, likely dominated by host flux). The right y-axis gives the apparent magnitude corrected for distance. Spectral epochs are given as vertical lines at the top of the plot.}
    \label{fig:multiband_lightcurve}
\end{figure*}

Photometry in $gri$ was obtained at the position of \xdos using the ZTF forced-photometry service\footnote{\href{https://ztfweb.ipac.caltech.edu/cgi-bin/requestForcedPhotometry.cgi}{{https://ztfweb.ipac.caltech.edu/cgi-bin/requestForcedPhotometry.cgi}}} \citep{Masci2019}. Quality checks are performed on the raw data and detections are chosen based on a signal-to-noise ratio (S/N) of S/N $\geq$ 3. Photometry from the ATLAS forced-photometry server\footnote{\href{https://fallingstar-data.com/forcedphot/}{https://fallingstar-data.com/forcedphot/}} \citep{Tonry2018,Smith2020,Shingles2021a} at the position of \xdos were obtained in the ATLAS $c$ and $o$ filters. We compute the weighted average of the fluxes of the observations on nightly cadence. We perform a quality cut of 3$\sigma$ in the resulting flux of each night for both filters and then convert them to the AB magnitude system.  A single epoch of photometry with the AAlhambra Faint Object Spectrograph and Camera,  (ALFOSC) on the 2.56 m Nordic Optical Telescope (NOT) was obtained in $gr$ and three epochs from the Liverpool Telescope (LT) with the optical imaging component of IO (Infrared-Optical) instrument was obtained in $gri$. J-band photometry was obtained with the Palomar Gattini-IR (PGIR) telescope shortly after \xdos was discovered and data were reduced following \citealt{De2020}. No significant activity/outbursts are detected in the pre-discovery images extending back to the beginning of the ZTF observing (circa 2018) to a limit of -14 mag. Our photometric dataset is given in Fig. \ref{fig:multiband_lightcurve} and photometric tables are available in the online material.

We observed the field \xdos with the Swift's onboard X-ray telescope XRT \citep{Burrows2005a} in photon-counting mode between 2022-02-23T16:47:35 (MJD: 59633.699) and 2022-04-16T12:19:33 (MJD: 59685.5136). We analysed all photon-counting data with the online tools of the UK Swift\ team that use the methods described by \citealt{Evans2007a, Evans2009a, Evans2020a} and the software package HEAsoft\footnote{\href{https://heasarc.gsfc.nasa.gov/lheasoft}{https://heasarc.gsfc.nasa.gov/lheasoft}}.

An image constructed from all XRT data reveals a fading source at RA, Decl. (J2000) 11:48:06.72, $-$12:38:45.0 with an uncertainty of 5.3 arcsec (radius, 90\% confidence), using the online tool of the UK Swift team  \footnote{ \href{http://www.swift.ac.uk/user_objects/}{http://www.swift.ac.uk/user\_objects/} }  \citep{Evans2007a, Evans2009a}. Using the dynamic binning method of the swift\ online tools, we obtain two detections summarised in Table \ref{tab:xrt}. The number of counts is too low to robustly constrain the X-ray spectrum. To convert the count-rate to a flux, we assume an absorbed power-law with a photon index of 2 where the absorption components is set to the Milky-Way value of $N({\rm H,Gal})=2.74\times10^{20}~\rm cm^{-2}$ from \citep{HI4PI2016a}. Using these spectral parameters, we derive count rate to flux conversion factor of $3.75\times10^{-11}~\rm erg\,s^{-1}\,cm^{-2}/\left(ct\,s^{-1}\right)$ to infer the unabsorbed flux between 0.3 and 10 keV. The converted count rates are summarised in Table \ref{tab:xrt}, and we will further discuss the X-ray luminosity in Sect. \ref{ssec:coronal}.

\begin{table}
    \caption{Log of XRT observations\label{tab:xrt}}
      \begin{tabular}{cccc}
      \toprule
      MJD & Phase  & Count rate               & $F~(0.3-10~\rm keV)$\\
      & (days) & ($10^{-3}~{\rm s}^{-1}$) & ($10^{-12}~\rm erg\,s^{-1}\,cm^{-2}$)\\
      \midrule
      $59633.9^{+26.1}_{-0.2}$& +113  & $0.011 \pm 0.003$ &$ 0.41 \pm 0.10$\\
      $59681.4^{+4.3}_{-13.5}$& +161 & $0.003 \pm 0.001$ &$ 0.12 \pm 0.05$\\
\bottomrule
\end{tabular}
\tablefoot{The flux reports the brightness after accounting for absorption in the Milky Way.}
\end{table}

Due to its proximity at only 78~Mpc,  and the fact that  \xdos is a slow evolving luminous transient, fading by only 4.5~mags during the 1.5 year (18 month) dataset presented in this work, this SN can potentially be followed-up for years.  Figure~\ref{fig:multiband_lightcurve} illustrates the entire multi-band observations of \xdos covering from discovery until solar conjunction in mid-2023. Spectroscopic follow-up began in February 2021, and is discussed further in Sect.~\ref{ssec:obsdata_spectra}.

\begin{figure}
    \centering
    \includegraphics[width= \columnwidth]{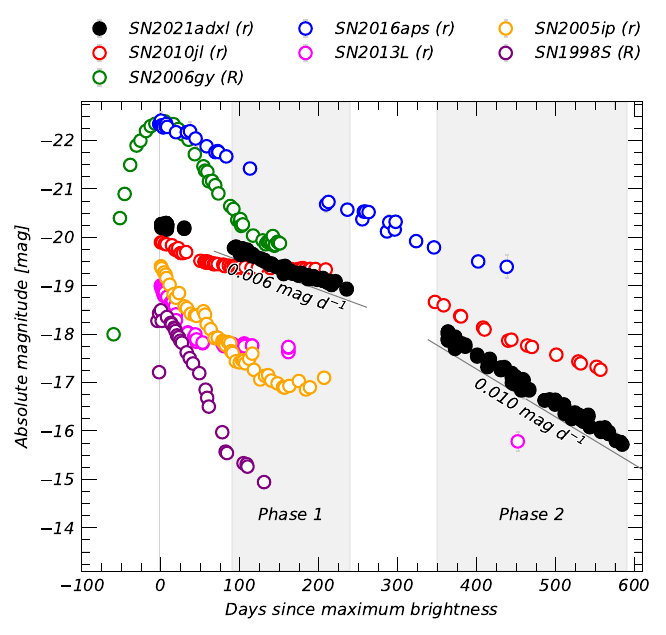}
    \caption{$r/R$-band lightcurve comparison of luminous interacting SNe including \xdos. Each transient is corrected for distance and MW extinction. The regions marked as Phase 1 and Phase 2 correspond to the phases in the post-peak evolution where \xdos shows unique decline rates. }
    \label{fig:lightcurve_comparision}
\end{figure}

The lightcurve of \xdos shows two distinct phases (referred to as Phase 1 and Phase 2) in its post-peak evolution. The peak itself is poorly sampled and it is unclear if Phase 1 extends to the earliest detections. Approximately 100 days post discovery, \xdos shows smooth decline of $\sim0.6$ \magperhundredday, declining faster in the bluer bands, as is typical for Type IIn SNe \citep[see Fig.~10 in][]{Nyholm2020}. After the mid-2022 solar conjunction, \xdos begins to decline at a slightly faster rate of 1 \magperhundredday.

The rise to peak for \xdos was not observed due to solar conjunction in late-2021. An important aspect of Type IIn SNe is their rise time to maximum light, as there is a strong correlation between rise time and CSM/ejecta properties \citep[e.g.][]{Suzuki2016,Nyholm2020}. The inset of Fig.~\ref{fig:multiband_lightcurve} shows the Gaia G-band observations of the site of \xdos covering the past 9 years. Prior to \xdos, there is no significant variability seen at the explosion site. Similarity, no precursor signal is seen in forced-photometry measurements of ZTF, and ATLAS observations.

\begin{figure}
    \centering
    \includegraphics[width= \columnwidth]{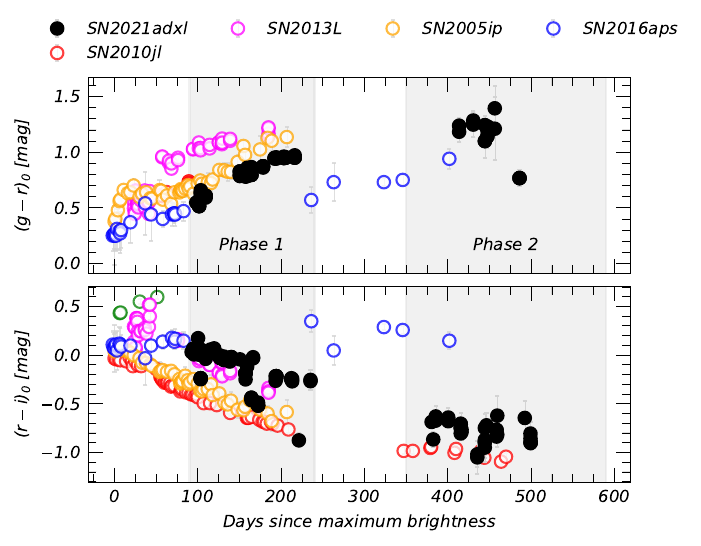}
    \caption{Color comparison of bright Type IIn SNe including \xdos. Each transient has been corrected for MW extinction.  }
    \label{fig:color_comparision}
\end{figure}

\begin{figure}
    \centering
    \includegraphics[width= \columnwidth]{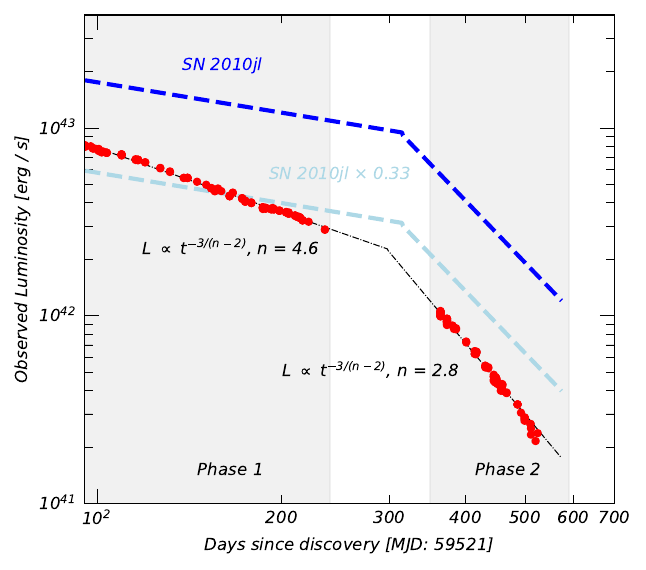}
    \caption{Bolometric light curve for \xdos.The black dashed lines show power law fits to the phase 1 and phase 2 light curve, which a distinct break at $\sim$ +300 days post discovery. The blue dashed line is the similar power law fits for SN~2010jl \citep{Fransson2014}, as well as the same fit scaled to match \xdos by eye in light blue. }
    \label{fig:luminosity}
\end{figure}

A pseudo-bolometric lightcurve is constructed using \software{Superbol}\footnote{\href{https://github.com/mnicholl/superbol}{https://github.com/mnicholl/superbol}} \citep{Nicholl2018} from the photometry presented in Fig.~\ref{fig:multiband_lightcurve}, and calibrated using the extinction and distance given in Sect.~\ref{sec:introduction}. The resulting lightcurve is shown in Fig.~\ref{fig:luminosity}. The bolometric light curve  can be accurately characterized by a power law decay from $\sim$ 0 - 300 days (with respect to discovery date), given by ${ \rm L \approx  7.81 \times 10^{42}  \times (t/100~days)^{-1.13} ~erg~s^{-1} }$ and a final steep decay given as ${ \rm L \approx  2.17 \times 10^{42}  \times (t / 300~days)^{-3.89}~erg~s^{-1} }$ after $\sim$ 300 days. As shown in Fig. \ref{fig:luminosity}, this is similar to SN~2010jl \citep{Fransson2014}, but \xdos is around 66\% less luminous.

Figure~\ref{fig:lightcurve_comparision} compares the $r/R$-band absolute light curve of \xdos with several transients with similar photometric appearance, most notably SN~2013L \citep{Andrews2017,Taddia2020} and SN~2010jl \citep{Fransson2014,Jencson2016}. Additionally, Fig.~\ref{fig:color_comparision} gives the extinction-corrected  color evolution for each transient. Qualitatively, \xdos, SN~2013L, and SN~2010jl share the same lightcurve morphology and color evolution. Post-peak, each of these transients light curves flattens after $\sim$ 100 days. \xdos shows a steeper decline compared to SN~2013L and SN~2010jl, with a slope similar to that seen in SN~2016aps \citep{Nicholl2020}, albeit $\sim$2~mags less luminous.

After $\sim$ 250 days, \xdos begins to decline faster, and the same evolution is also clearly seen for  SN~2010jl and SN~2013L, although with much poorer cadence for the latter. A similar trend was observed also for the Type IIn SN~2005ip, however for that SN the first flattening stage happened later, at approximately +200 days \citep{Stritzinger2012} and  the second drop-off  much later, at +3000 days \citep{Fox2020}. For contrast, the prototype Type IIn SN~1998S does not show any light-curve flattening, but declines rather quickly \citep{Schlegel1990,Mauerhan2012}.

As seen from Fig.~\ref{fig:lightcurve_comparision}, there is a spread in peak magnitudes for the transients. Under the assumption that the explosion mechanism for each transient is similar, e.g. core-collapse \citep{Bethe1990,Woosley1995,Smartt2009} or pulsational-pair instabilities \citep{Woosley2017}, the heterogeneity in brightness could be the result of different amounts of CSM and/or ejecta mass, although this assumes the explosion energy is similar for all Type IIns. We further discuss the ejecta and CSM properties of \xdos in Sect.~\ref{ssec:modelling}

\subsection{Spectral evolution}\label{ssec:obsdata_spectra}

\begin{figure}
    \centering
        \includegraphics[width= \columnwidth]{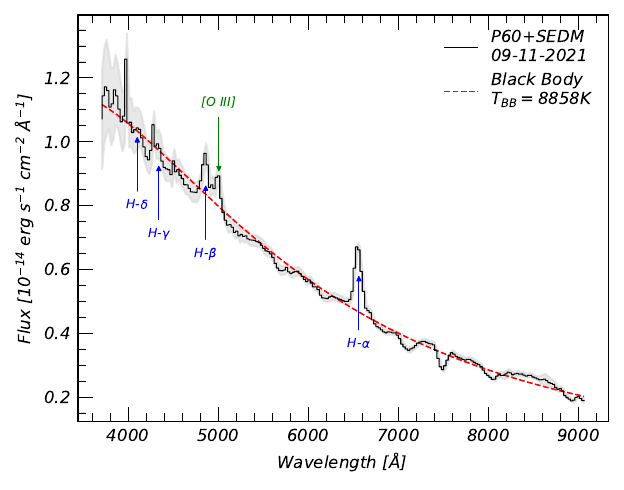}
    \caption{The earliest spectrum of \xdos from P60/SEDM taken on 2021-11-09. The mean flux is given in black, with 1$\sigma$ errors shaded in grey.}
    \label{fig:earliest_spectrum}
\end{figure}

\begin{figure}
    \centering
    \includegraphics[width= \columnwidth]{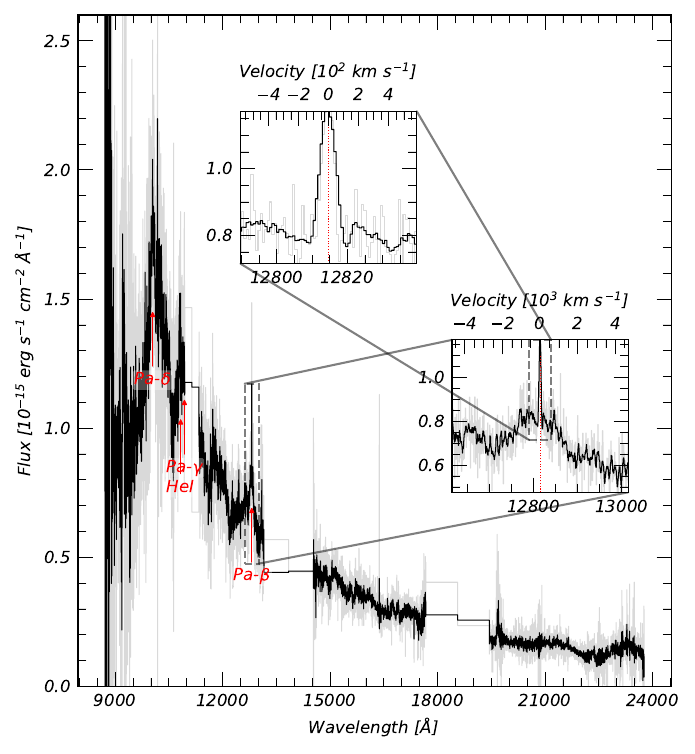}
    \caption{Classification NIR spectrum of \xdos from the Magellan/FIRE from 2022-02-01 \citep{De2022}. A close-up of the  Pa-$\beta$ line is given in the inset on the right, with a zoomed-in profile in the upper left. The spectrum smoothed with a Savgol filter is given in black, with the raw data given in grey.}
    \label{fig:FIRE_spectrum}
\end{figure}

Figure~\ref{fig:earliest_spectrum} shows the earliest spectrum of \xdos, taken around the(apparent) peak brightness. Although the spectrum shows some undulations, its overall appearance is consistent with a Type IIn SN, i.e a blue continuum with narrow Hydrogen emission features, with underlying host flux seen in [\ion{O}{III}] [\ion{O}{III}] $\lambda 5007$. \xdos was then classified as a Type IIn by \citealt{De2022} using the Magellan/Baade with the FIRE NIR instrument on 2022-02-01 and the classification spectrum is shown in Fig.~\ref{fig:FIRE_spectrum}.

Follow-up spectra were obtained using NOT+ALFOSC, the Low Resolution Imaging Spectrometer \citep[LRIS; ][]{Oke1995} on the 10 m Keck telescope, the Double Spectrograph (DBSP) on the Palomar 200-inch telescope (P200), and the Spectral Energy Distribution Machine \citep[SEDM;][]{Blagorodnova2018,Rigault2019} on the Palomar 60-inch (P60) telescope. Observations using the P60 and P200 were coordinated using the FRITZ data platform \citep{Walt2019,Coughlin2023}. The spectra were reduced in a standard manner, using \software{LPipe}  \citep{Perley2019}, \software{DBSP\_DRP} \citep{Mandigo2022} and \software{PypeIt} \citep{pypeit:zenodo,pypeit:joss_arXiv,pypeit:joss_pub}, for Keck/LRIS, P200/DBSP, and NOT/ALFOSC, respectively, and SEDM spectra were reduced using the automated pipeline \software{pysedm} \citep{Rigault2019,Kim2022}.

\begin{figure*}
    \centering
    \includegraphics[width= \textwidth]{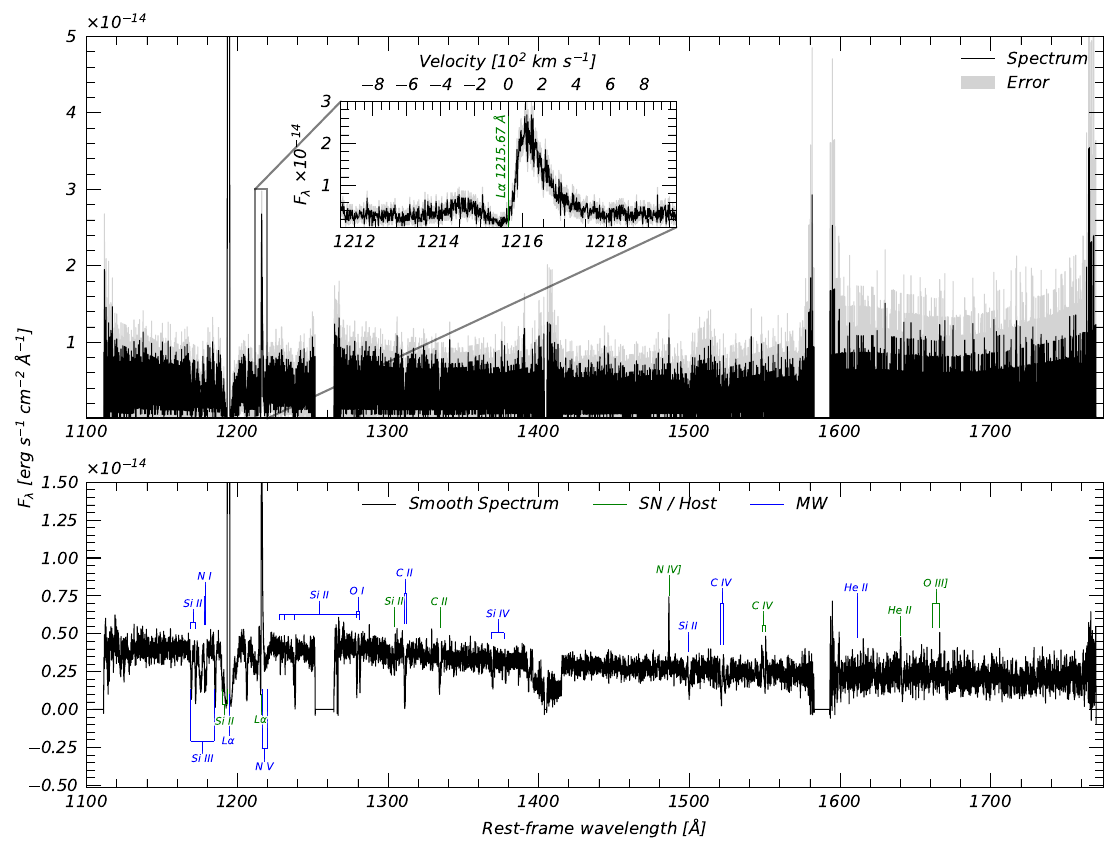}
    \caption{HST/COS UV spectrum of \xdos taken with the G130M (FUV) and G160M (NUV) grism  on 2022-04-11 (+158d). Upper panel shows the (error) spectrum given in (grey) black. Lower panel gives the science spectrum smoothed using a Savitzky–Golay (Savgol) filter with window size equal to 31 with a 2nd order polynomial. We include expected wavelengths for both Milky Way (MW) (blue) and \xdos plus its host (Green), although the detection of some of these lines are unclear due to the low S/N of the science spectrum.   }
    \label{fig:HST_COS}
\end{figure*}

A single UV spectrum was obtained using the Cosmic Origins Spectrograph (COS) onboard the Hubble Space Telescope (HST) on 2022-04-11 (Program ID: 16931, PI: Yan) and is presented in Fig.~\ref{fig:HST_COS}. This spectrum is low S/N (upper panel of Fig.~\ref{fig:HST_COS}), however when a smoothing filter is applied, we recover interstellar absorption bands \citep[see ][ and references therein]{Haser1998}, and several emission lines at the redshift of \xdos. Most features seen in the lower panel of Fig.~\ref{fig:HST_COS} likely arise due to interstellar absorption within the MW, although we make tentative detections of \ion{N}{iv}], \ion{O}{iii}], and \ion{He}{ii}, as well as \lalpha, at the redshift of \xdos.

\begin{figure}
    \centering
    \includegraphics[width= \columnwidth]{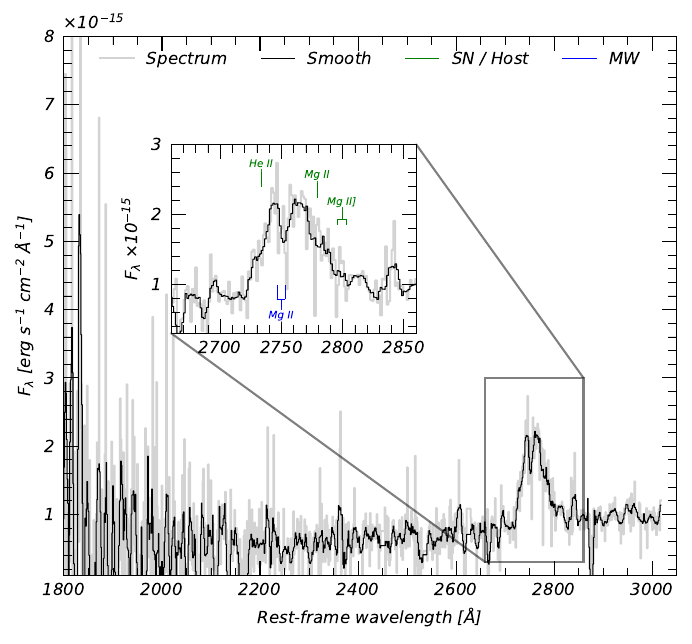}
    \caption{HST/STIS with the G230LB grism  on 2022-04-11 (+158d). The original science spectrum (grey) has been smoothed (black) using a Savgol filter with window size equal to 11 with a 2nd order polynomial. In the inset we show a close up of the broad feature around 2759$\AA$ and include expected wavelengths for both interstellar (blue) and \xdos (green) lines. }
    \label{fig:HST_STIS}
\end{figure}

A near UV spectrum was obtained on 2022-04-11 (Program ID: 16931, PI: Yan) using the Space Telescope Imaging Spectrograph (STIS) abroad HST and is presented in Fig. \ref{fig:HST_STIS}. The spectrum's appearance does show some continuum flux, likely from \xdos, with a single broad feature at $\sim2758\AA$. While this may be broad/blended \ion{Mg}{II} $\lambda\lambda2796, 2803$ (albeit offset by $F\sim$40$\AA$), it is also a possible a blend of several host/transient lines with superimposed interstellar lines, see inset of Fig. \ref{fig:HST_STIS}. The UV appearance of \xdos and its host is further discussed in Sect. \ref{ssec:green_pea}.

We obtained four medium-resolution spectra with the \xshooter instrument \citep{Vernet2011a} between 2022-02-26 and 2023-02-26 (observing program 108.2262; PI. R. Lunnan). All observations were performed in nodding mode and with 1\farcs0 / 0\farcs9 / 0\farcs9 wide slits (UVB / VIS / NIR). The first four epochs covered the full spectral range from 3000 to 24,800~\AA. The integration times were varied between 2400 and 3700s. The data were reduced following \citealt{Selsing2019a}. In brief, we first removed cosmic-rays with the tool \software{astroscrappy}\footnote{\href{https://github.com/astropy/astroscrappy}{https://github.com/astropy/astroscrappy}}, which is based on the cosmic-ray removal algorithm by \citealt{vanDokkum2001a}. Afterwards, the data were processed with the \xshooter pipeline v3.3.5 and the ESO workflow engine ESOReflex \citep{Goldoni2006a, Modigliani2010a}. The UVB and VIS-arm data were reduced in stare mode to boost the S/N. In the background limited case, this can increase the S/N by a factor of $\sqrt{2}$ compared to the standard nodding mode reduction (in the background limited case). The individual rectified, wavelength- and flux-calibrated two-dimensional spectra files were co-added using tools developed by J. Selsing\footnote{\href{https://github.com/jselsing/XSGRB_reduction_scripts}{https://github.com/jselsing/XSGRB\_reduction\_scripts}}. The NIR data were reduced in nodding mode to ensure a good sky-line subtraction. In the third step, we extracted the one-dimensional spectra of each arm in a statistically optimal way using tools by J. Selsing. Finally, the wavelength calibration of all spectra were corrected for barycentric motion. The spectra of the individual arms were stitched by averaging the overlap regions. Note, to study the host galaxy, we reduced the data in nodding mode.

\begin{figure*}
    \centering
    \includegraphics[width= \textwidth]{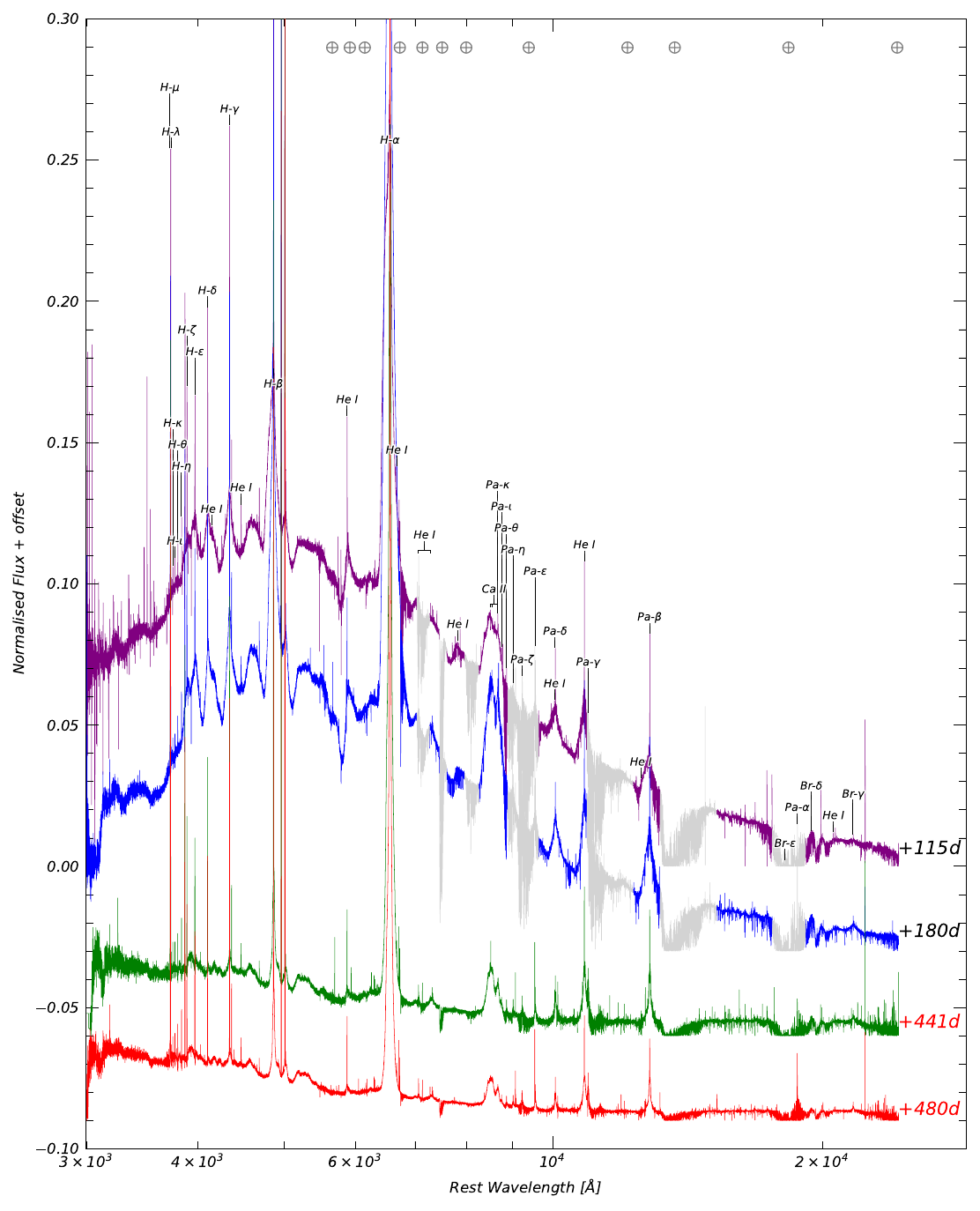}
    \caption{VLT + \xshooter for \xdos taken at four different epochs. Each spectrum has been normalised with respect to H$\alpha$ and offset for clarity. Emission lines expected from the transient (black)  are denoted by the vertical lines. Telluric absorption bands arising from atmospheric \rm{O$_2$} and \rm{H$_2$O} are denoted by the {\rm$\oplus$} symbol. We shade in regions of high noise levels from our first two spectra for visual clarity. }
    \label{fig:X-shooter_spectra}
\end{figure*}

Figure~\ref{fig:X-shooter_spectra} presents four epochs of spectroscopy taken with the VLT and the \xshooter medium resolution spectrograph. All spectra was observed at parallactic angle with the UV, optical and IR regions stitched using a chi-2 minimization and flux calibrated top photometry taken on the same night. Two spectra were obtained during Phase 1 of the lightcurve, and two during Phase 2. The spectra show contributions from both transient light, seen by the broader emission profiles, as well as flux from the underlying host. At the time of writing, \xdos still dominates the flux at the location, and templates are currently unobtainable in order to remove narrow emission components from the host.A log of the available spectra is provided in Table~\ref{tab:observation_log} and the epochs of spectroscopy are indicated in Fig.~\ref{fig:multiband_lightcurve} as the vertical lines. All wavelength reports are given in rest-frame unless otherwise stated.

\begin{table*}
\centering
\caption{Spectropscopy log for \xdos. Phase with reference to discovery date [MJD 59521]}
\label{tab:observation_log}
\begin{tabular}{ccccccc}
\toprule
Date & MJD & Phase (d) & Telescope & Instrument & Resolution ($\lambda / \Delta \lambda $) & Exposure Time (s) \\
\midrule
2021-11-09 & 59527 & +6 & P60 & SEDM IFU & 100 & 2400 \\
2022-02-02 & 59612 & +91 & P200 & DBSP & 4000 & 300 \\
2022-02-03 & 59613 & +92 & P60 & SEDM IFU & 100 & 2400 \\
2022-02-04 & 59614 & +93 & P60 & SEDM IFU & 100 & 2400 \\
2022-02-05 & 59615 & +94 & Keck & LRIS  & 300-5000 & 2400 \\
2022-02-05 & 59615 & +94 & P60 & SEDM IFU & 100 & 2400 \\
2022-02-06 & 59616 & +95 & P60 & SEDM IFU & 100 & 2400 \\
2022-02-07 & 59617 & +96 & P60 & SEDM IFU & 100 & 2400 \\
2022-02-09 & 59619 & +98 & P60 & SEDM IFU & 100 & 2400 \\
2022-02-10 & 59620 & +99 & P60 & SEDM IFU & 100 & 2400 \\
2022-02-11 & 59621 & +100 & P60 & SEDM IFU & 100 & 2400 \\
2022-02-14 & 59624 & +103 & NOT & ALFOSC+Grism4 & 360 & 1800 \\
2022-02-20 & 59630 & +109 & P60 & SEDM IFU & 100 & 2400 \\
2022-02-26 & 59636 & +115 & VLT & X-shooter & 5400/8900/5600 & 2400/2458/2400 \\
2022-02-26 & 59636 & +115 & P60 & SEDM IFU & 100 & 2400 \\
2022-03-02 & 59640 & +119 & P60 & SEDM IFU & 100 & 2160 \\
2022-03-09 & 59647 & +126 & P60 & SEDM IFU & 100 & 2400 \\
2022-03-14 & 59652 & +131 & P60 & SEDM IFU & 100 & 2400 \\
2022-03-21 & 59659 & +138 & P60 & SEDM IFU & 100 & 2400 \\
2022-03-21 & 59659 & +138 & P60 & SEDM IFU & 100 & 2400 \\
2022-03-23 & 59661 & +140 & P60 & SEDM IFU & 100 & 2400 \\
2022-03-23 & 59661 & +140 & P60 & SEDM IFU & 100 & 2400 \\
2022-03-28 & 59666 & +145 & P60 & SEDM IFU & 100 & 2400 \\
2022-03-30 & 59668 & +147 & NOT & ALFOSC+Grism4 & 360 & 1800 \\
2022-04-01 & 59670 & +149 & P60 & SEDM IFU & 100 & 2400 \\
2022-04-11 & 59680 & +159 & HST & COS/STIS & 2000-3000 & 1911 \\
2022-04-11 & 59680 & +159 & P60 & SEDM IFU & 100 & 2400 \\
2022-04-18 & 59687 & +166 & P60 & SEDM IFU & 100 & 2160 \\
2022-04-26 & 59695 & +174 & NOT & ALFOSC+ Grism4 & 360 & 1800 \\
2022-05-02 & 59701 & +180 & VLT & X-shooter & 5400/8900/5600 & 3600 \\
2022-05-15 & 59714 & +193 & P60 & SEDM IFU & 100 & 2160 \\
2022-05-27 & 59726 & +205 & NOT & ALFOSC+Grism4 & 360 & 1800 \\
2022-06-12 & 59742 & +221 & P60 & SEDM IFU & 100 & 2160 \\
2022-11-09 & 59892 & +371 & P60 & SEDM IFU & 100 & 2400 \\
2022-11-25 & 59908 & +387 & NOT & ALFOSC+Grism4 & 360 & 1200 \\
2022-12-08 & 59921 & +400 & P60 & SEDM IFU & 100 & 1800 \\
2023-01-18 & 59962 & +441 & VLT & X-shooter & 5400/8900/5600 & 1800/1860/1800 \\
2023-02-26 & 60001 & +480 & VLT & X-shooter & 5400/8900/5600 & 3600/3716/3600 \\
2023-03-23 & 60026 & +505 & NOT & ALFOSC+Grism4 & 360 & 3300 \\
2023-04-16 & 60050 & +529 & NOT & ALFOSC+Grism4 & 360 & 3300 \\
\bottomrule
\end{tabular}
\end{table*} 

\subsection{Hydrogen spectral evolution}\label{ssec:hydrogen_evolution}

The most prominent and informative features in the spectra of Type IIn SNe is the Balmer series, and specifically the \halpha profile \citep[e.g. ][]{Roming2012,Ransome2021,Pursiainen2022}. The \halpha emission line has been used to constrain the progenitor wind velocity \citep{Kankare2012,Andrews2017,Chugai2019}, explosion geometry \citep{Hoffman2008,Andrews2018,Pursiainen2022}, and can potentially indicate late-time interaction \citep{Silverman2013,Yan2015}.

Type IIn SNe typically develop some asymmetries in their \halpha emission line profiles in the post peak evolution \citep{Trundle2009,Fransson2014,Andrews2017,Andrews2019,Moriya2020,Taddia2020,Pursiainen2022}, and there are also cases where multiple absorption troughs are observed \citep{Gutierrez2017,Brennan2022}. As noticed by \citet{De2022}, the NIR classification spectrum for \xdos shows strong signs of hydrogen and helium emission. The insets of Fig.~\ref{fig:FIRE_spectrum} give a close-up view of the Pa-$\beta$ line, which is the best isolated hydrogen emission feature. In agreement with the initial classification, we resolve two apparent absorption features for the Pa-$\beta$ line, one with a trough centered around $\sim$3500\kms (right inset) and the second at $\sim$200\kms (upper left inset). Although the latter has low S/N, a consistent absorption feature is seen also in other Paschen lines. 

We observe a P-Cygni profile at the approximate wavelength of \lalpha in Fig. \ref{fig:HST_COS}. The emission of \lalpha is redshifted by $\sim~100$~\kms, with a blueshifted absorption centered at $\sim~-100$~\kms (with respect to rest wavelength). Although the appearance of \lalpha for \xdos is in contrast to the profile seen for SN~2010jl \citep{Fransson2014}, with \xdos showing a P-Cygni profile, and SN~2010jl showing a purely absorption profile. This redshifted profile for \xdos may be a result of electron scattering \citep{Huang2018}. As this spectrum is obtained $\sim5$ months post discovery, the P-Cygni absorption likely originates from the pre-supernova stellar wind, located at large distances from the ejecta-CSM interface. This is consistent with the absorption seen earlier in Fig. \ref{fig:FIRE_spectrum}. This would also suggest that \xdos has a very extended, dense circumstellar environment, which is likely correlated with the slow evolution observed in Fig. \ref{fig:multiband_lightcurve}.  However, we can speculate that the strange \lalpha emission profile is a result of the host environment itself. We discuss this further in Sect. \ref{ssec:host}

\begin{figure}
    \centering
    \includegraphics[width= \columnwidth]{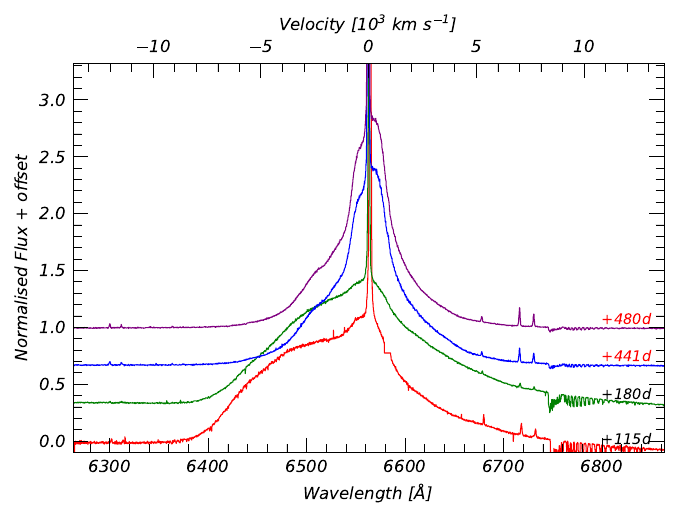}
    \caption{Evolution of the \halpha profile from the VLT/\xshooter observations. A pseudo-continuum has been removed from each profile, and the spectra are normalised and offset for clarity. Phase given in grey (red) represents spectra obtained during (after) Phase 1. A noise spike at 6852$\AA$ was manually removed from the +115~d spectrum.}
    \label{fig:X-shooter_halpha_evolution}
\end{figure}

Figure~\ref{fig:X-shooter_halpha_evolution} highlights the evolution of the \halpha profile from the VLT/\xshooter observations. The \halpha profile consists of multiple distinct components, giving rise to an overall asymmetric appearance. A prominent blue-shifted emission component is observed in \halpha and in other Balmer lines. This feature is not obviously seen in other emission features, for example in \ion{He}{I} or \ion{Fe}{II}, which suggests that the physical parameters forming this component are unique to a hydrogen-rich shell of  material. This broad component does not extend to redder wavelengths, but the rather smooth line wings are indicative of electron scattering \citep{Huang2018}. From the earliest \xshooter spectra, a third broad component can be resolved around rest wavelength with a superimposed narrow emission line likely arising from slow-moving CSM, an underlying \ion{H}{II} region, or a combination of both.

\begin{figure}
    \centering
    \includegraphics[width= \columnwidth]{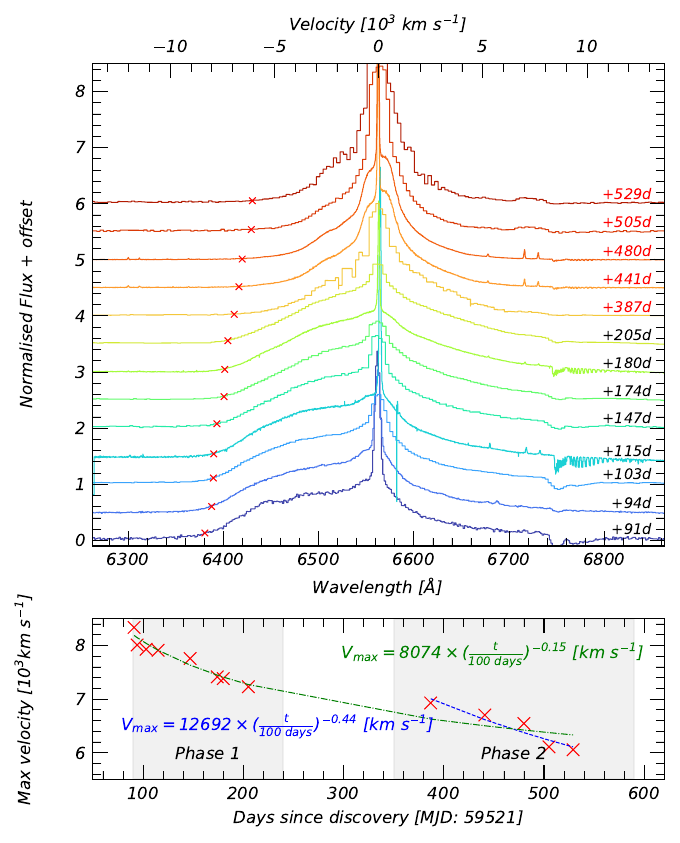}
    \caption{Same as Fig.~\ref{fig:X-shooter_halpha_evolution} but including the complete optical spectral dataset (excluding those from the P60/SEDM). A red cross on each profile denotes the maximum velocity of the blue excess for each spectrum. The lower panel gives the velocity of the bluest edge of the \halpha profile, marked by the red cross in the upper panel.}
    \label{fig:halpha_evolution}
\end{figure}

As \xdos evolves, we see a smooth evolution of the \halpha profile, most notably of the blueward component. The upper panel of Fig.~\ref{fig:halpha_evolution} shows the complete optical spectral dataset (excluding the low-resolution P60+SEDM spectra). We mark the point where the flux of \halpha exceeds 10\% of the pseudo-continuum with a red cross. This component shows a smooth evolution during Phase 1, see lower panel of Fig.~\ref{fig:halpha_evolution}. During Phase 2 it is unclear if the \halpha velocity continues to follow this trend as it apparently shows a quicker decline in velocity.

\begin{figure}
    \centering
    \includegraphics[width= \columnwidth]{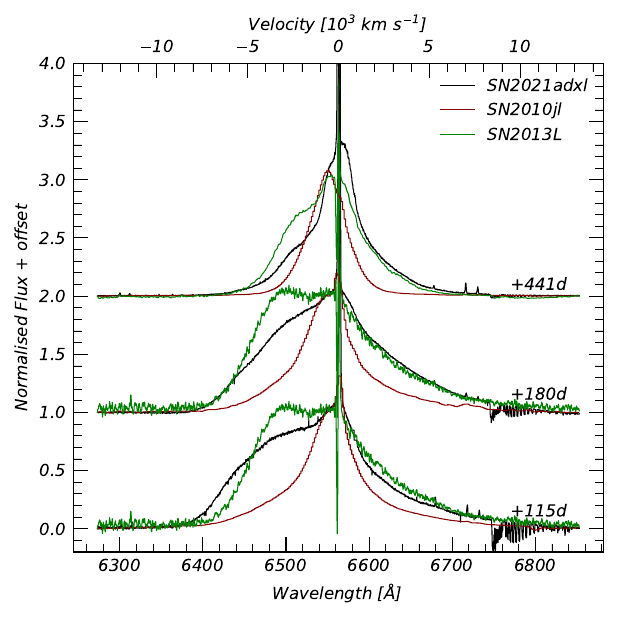}
    \caption{Comparison of the \halpha profile from \xdos, SN~2010jl \citep{Fransson2014}, and SN~2013L \citep{Andrews2017,Taddia2020}. Each transient shows a blue shoulder atop their respective \halpha profile, as well as broad wings. Each profile has been normalised and offset for clarity.}
    \label{fig:compare_halpha}
\end{figure}

Compared to photometrically similar objects, \xdos shows qualitative similarities to the Type IIn SNe 2010jl \citep{Fransson2014}, 2013L \citep{Andrews2017,Taddia2020} as well as SN~2005ip \citep{Fox2020}. Figure~\ref{fig:compare_halpha} compares the \halpha profile of \xdos, SN~2010jl, and SN~2013L. Each transient shows a smooth redward wing, likely a result of electron scattering, as well as a blueward excess that evolves with time.  It should be noted that while these three transients are qualitatively similar photometrically and spectroscopically, the photometric decline times and the emission line amplitude and offset are different among these events, and this may be due to differences in mass, velocity, and distribution of the emitting material. This might also be reflected in the different lightcurves, see Fig.~\ref{fig:lightcurve_comparision}. Insights can be gained by noting the +115d spectrum of \xdos compared to a spectrum of SN~2010jl at a similar epoch (+94d). \xdos shows flux out to 8000\kms while SN~2010jl only shows a slight blue excess out to 2000\kms, however this excess lines up quite well with the central wavelengths of the \halpha profile at this time. Whereas the \halpha profile of SN~2013L at this time matches the red wing of \xdos very well, and shows a much more dramatic flat-topped blue excess that persists for almost 4 years post peak \citep{Andrews2017}. Comparing the photometric evolution, SN~2013L shows a rapid decline before flattening, whereas SN~2010jl is more similar to \xdos, albeit declining at a slower rate.

\subsection{Helium spectral evolution}\label{ssec:helium_evolution}

\begin{figure}
    \centering
    \includegraphics[width= \columnwidth]{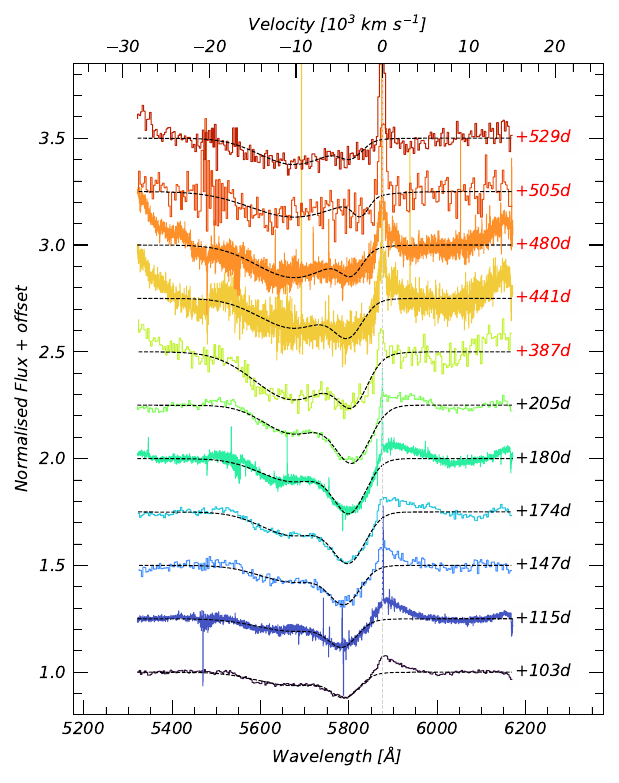}
    \caption{Evolution of \ion{He}{I} $\lambda 5876$ line. Each profile is continuum-subtracted. A double Gaussian absorption profile is fitted to the blue side of each spectrum, and is given by the dashed black line. We do not attempt to include the components in emission but note that the emission feature at rest wavelength narrows with time.}
    \label{fig:HeI5876_evolution}
\end{figure}

Aside from hydrogen, the spectrum of \xdos is dominated by strong Helium emission features, such as \ion{He}{I} $\lambda4471$, $\lambda 5876$, $\lambda 7065$ and $\lambda 10830$, and possibly \ion{He}{I} $\lambda2773$ (see Fig. \ref{fig:HST_STIS}). In stark contrast to the \halpha line, the profile of \ion{He}{I} $\lambda 5876$ resembles a P-Cygni profile. Figure~\ref{fig:HeI5876_evolution} provides a cut-out around \ion{He}{I} $\lambda 5876$, highlighting the strange P-Cygni-like profile with an absorption feature at $\sim$4000\kms.

Rather than the typical single absorption \citep[e.g. ][]{Israelian1999}, \ion{He}{I} $\lambda 5876$ might show a second high velocity (HV) feature, extending out to $\sim$18000\kms. It is unclear whether this HV absorption is indeed associated with \ion{He}{I} or a feature at bluer wavelength e.g. the \ion{Na}{I} D doublet $\lambda\lambda$5890, 5896, which is blended with the \ion{He}{I} $\lambda 5876$ line. If this is this case, a large amount of sodium-rich material must remain optically think more than 500 days post peak, and continue moving at very high velocities. In comparison to SN~2010jl, this double troughed absorption profile is not immediately obvious. In particular, the \ion{He}{I} $\lambda 5876$ observed for SN~2010jl resembles the \halpha profile e.g. see Fig. \ref{ssec:coronal}

An alternative suggestion is that this HV material is associated with the \ion{He}{I} $\lambda 5876$ line, although this feature is not clearly observed in other \ion{He}{I} lines, due to severe line blending and/or tellurics bands, see Fig. \ref{fig:X-shooter_spectra}. Seeing both absorption components would require  an asymmetric explosion/CSM in order to see both the low and high velocity component, although maintaining a P-Cygni -like absorption profile after 500 days is also difficult to explain.

\begin{figure*}
    \centering
    \includegraphics[width= \textwidth]{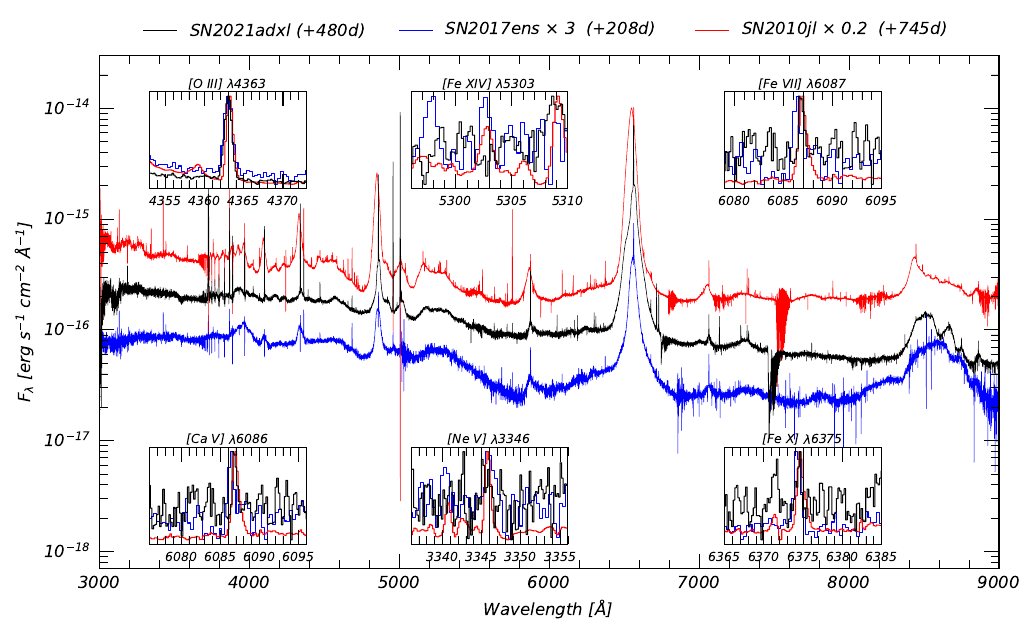}
    \caption{ Comparison of \xdos to the  Type IIn SN~2010jl \citep{Fransson2014}, as well as the transitional Type Ic-BL/IIn SN 2017ens \citep{Chen2018}. The insets highlight the wavelength range around known coronal emission lines. Each inset has been normalised to the peak of each respective emission line. Interacting transients are expected to produced a significant X-ray flux, which can be manifested in coronal line emission. Although these observations are rare (due to the late of high resolution spectra at late time), high-ionisation emission feature may be common in late time spectra of interacting SNe. }
    \label{fig:SN2021adxl_coronal}
\end{figure*}

\subsection{X-ray emission from Interacting SNe}\label{ssec:coronal}

When the progenitor of \xdos exploded, a shock wave was formed due the collision of the fast moving ejecta and the slower moving CSM. X-ray emission, if detected, can be used to diagnose the shock wave interaction with the circumstellar medium \citep{Chevalier2003,Dwarkadas2019}. We reported the detection of a decaying X-ray flux at the position of \xdos (see Table. \ref{tab:xrt}). This evolving X-ray is likely associated with \xdos and not the underlying source itself. We find the soft X-ray flux decreases from $\sim 3 \times 10^{40}~{\rm ergs~s^{-1}}$ at +113d to  $\sim 0.9 \times 10^{40}~{\rm ergs~s^{-1}}$ at +160d. This flux is roughly a factor of 10 less than SN~2010jl \citep{Chandra2015a}. However, meaningful comparison is moot due to the restricted X-ray cadence.

High ionization lines have been observed the optical spectra of nearby active galactic nuclei (AGNs) \citep{Oke1968,Appenzeller1988,Lamperti2017}, however they are rarely observed in SNe \citep{Turatto1993,Smith2009}, likely due to the lack of high to medium resolution spectra \citep{Groningsson2006,Komossa2009,Smith2009b,Pastorello2015,Fransson2014}. These so-called ``coronal lines'', first observed in the solar corona, arise from collisionally excited forbidden fine-structure transitions in highly ionized species such as Fe VI and \ion{Fe}{X}. 

Figure \ref{fig:SN2021adxl_coronal} shows our final \xshooter spectrum taken on 2023-02-26 with insets showing the location of several coronal emission lines. In terms of explosive transients, coronal lines have been interpreted in terms of dissipation of the energy of a shock wave in the circumstellar envelope. If collisional ionization is the dominant mechanism, then the presence of these lines implies a temperature of the emitting material of $10^5-10^6$ K \citep{Bryans2009}  and pre-shock ionization of the circumstellar medium by X-ray emission. For \xdos, we do not detect [\ion{Fe}{XIV}] $\lambda5303$ (see inset in Fig. \ref{fig:SN2021adxl_coronal}), while [\ion{Fe}{X}] $\lambda6375$ is strong meaning the CSM temperature is likely  $\lesssim2\times10^6$~K \citep{Jordan1969,Turatto1993}.

\section{Host environment of \xdos}\label{ssec:host}

\begin{figure}
    \centering
    \includegraphics[width= \columnwidth]{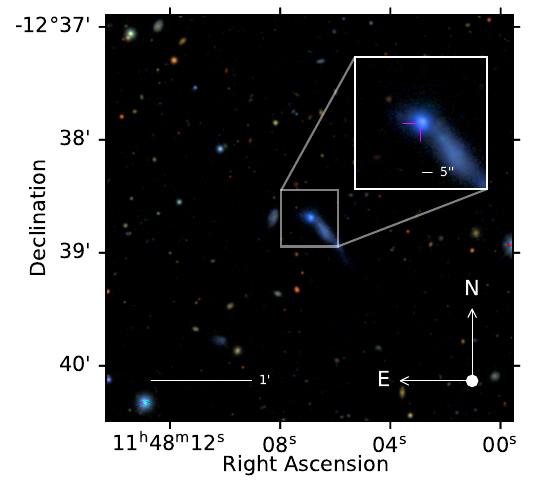}
    \caption{Composite ($gri$) color image of the host galaxy of \xdos obtained from the DESI Legacy Surveys \citep{Dey2019}.  \xdos (not visible in this image) exploded in the North-East part of its host, in a bright star-forming area.  The inset provides a zoom-in of the explosion site, where the position of \xdos  is marked by the crosshair. }
    \label{fig:finder_host}
\end{figure}

\xdos exploded in a blue luminous host, composed of a bright star-forming environment in the north-east, and an extended diffuse tail to the south-west (see Fig. \ref{fig:finder_host}).  This unique morphology is coined a ``tadpole'' \citep[][]{Elmegreen2012}. The blue compact northern region implies a high star-formation rate (SFR) \citep{Botticella2012,Binder2018}, and is a likely location for massive stars and a place where CCSNe occur.

To further understand the host environment of \xdos, the software package \software{Prospector} version 1.1 \citep{Johnson2021} was used to analyse the spectral energy distribution from the host.

We retrieved science-ready coadded images from the \textit{Galaxy Evolution Explorer} (GALEX) general release 6/7 \citep{Martin2005a}, the Panoramic Survey Telescope and Rapid Response System (Pan-STARRS, PS1) DR1 \citep{Chambers2016a} and images from the Wide-Field Infrared Survey Explorer \citep[WISE;][]{Wright2010a} processed by \citealt{Lang2014a}. We measured the brightness of the host using LAMBDAR\footnote{\href{https://github.com/AngusWright/LAMBDAR}{https://github.com/AngusWright/LAMBDAR}} \citep[Lambda Adaptive Multi-Band Deblending Algorithm in R;][]{Wright2016a} and the methods described in \citealt{Schulze2021a}. We also extract the photometry of the isolated star-forming region, in which the SN occurred.Table \ref{tab:host_phot} the measurements in the different bands, and Table \ref{tab:knot_phot} focuses on the bright star forming knot.

\begin{table}
\centering
\begin{tabular}{l c}
\hline
Instrument/Filter & Brightness \\
\hline
GALEX/FUV & $17.71 \pm 0.07$ \\
GALEX/NUV & $17.75 \pm 0.03$ \\
PS1/g & $17.26 \pm 0.05$ \\
PS1/r & $17.21 \pm 0.06$ \\
PS1/i & $17.11 \pm 0.04$ \\
PS1/z & $17.12 \pm 0.34$ \\
PS1/y & $17.19 \pm 0.05$ \\
WISE/W1 & $18.13 \pm 0.10$ \\
WISE/W2 & $18.50 \pm 0.20$ \\
\hline
\end{tabular}
\caption{Photometric measurements of the host of \xdos.  All magnitudes are in the AB system and not corrected for extinction.}
\label{tab:host_phot}
\end{table}

\begin{table}
\centering
\begin{tabular}{l c}
\hline
Instrument/Filter & Brightness \\
\hline
GALEX/FUV & $18.00 \pm 0.08$ \\
GALEX/NUV & $18.11 \pm 0.05$ \\
PS1/g & $17.97 \pm 0.01$ \\
PS1/r & $18.14 \pm 0.01$ \\
PS1/i & $17.93 \pm 0.01$ \\
PS1/z & $17.95 \pm 0.03$ \\
PS1/y & $18.12 \pm 0.02$ \\
\hline
\end{tabular}
\caption{Photometry of the isolated star-forming region where \xdos exploded. All magnitudes are in the AB system and not corrected for extinction.}
\label{tab:knot_phot}
\end{table}

\begin{figure}
    \centering
    \includegraphics[width= \columnwidth]{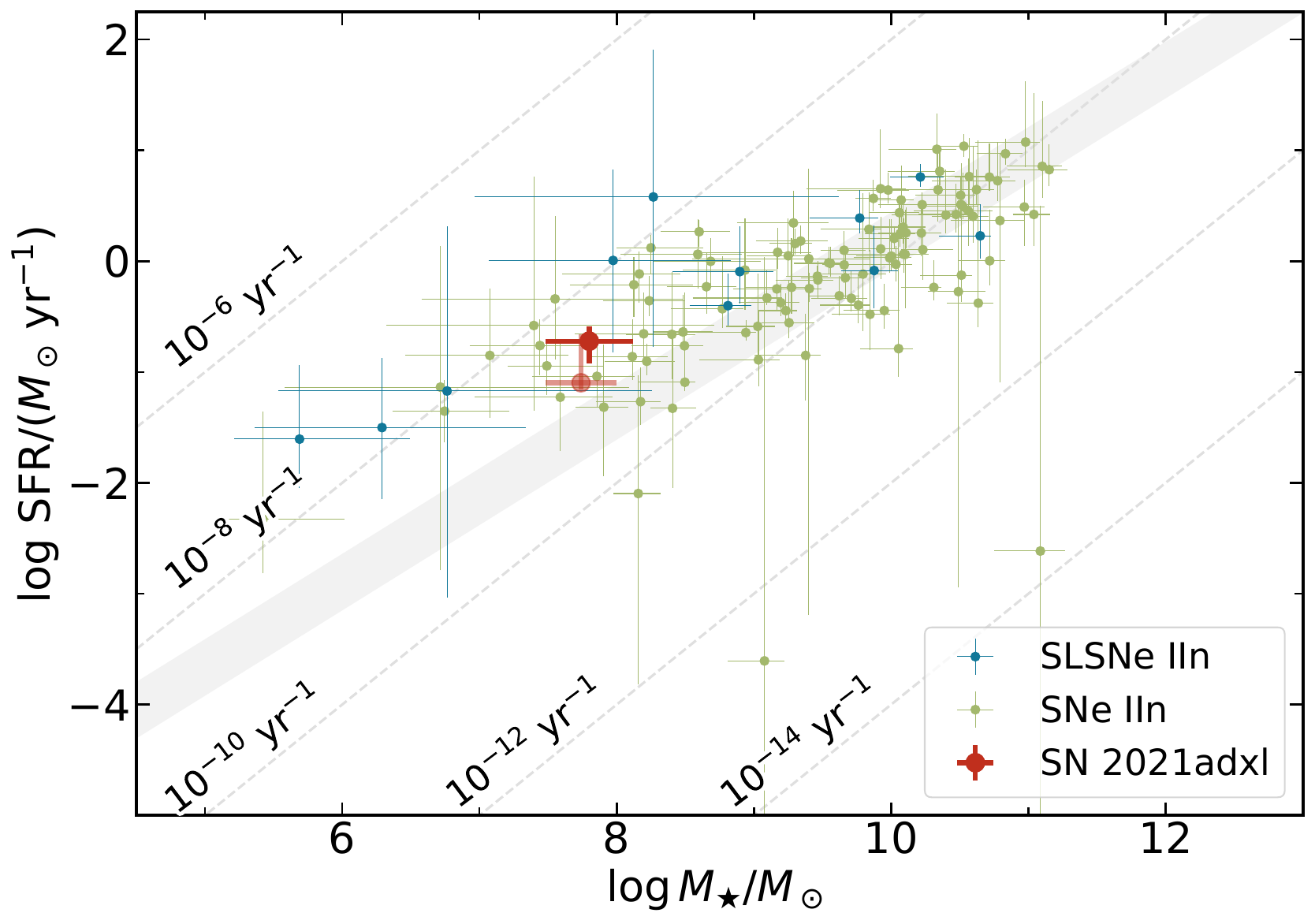}
    \caption{The star-formation rate and stellar mass of the host galaxy of \xdos (dark red) and of the bright star-forming region where the SN exploded (light shade) in the context of SN-IIn and SLSN-IIn host galaxies from the PTF survey \citep{Schulze2021a}. The host galaxy of \xdos lies in the expected parameter space of SN-IIn and SLSN-IIn host galaxies but in the lower half of the mass and SFR distributions. Its specific star-formation rate (SFR / mass) is slightly larger than the typical star-forming galaxies (grey band) but lower than for an average SLSN host galaxy.}
    \label{fig:SFR-mass}
\end{figure}

\begin{figure}
  \centering
  \begin{subfigure}{\columnwidth}
    \centering
    \includegraphics[width=0.8\columnwidth]{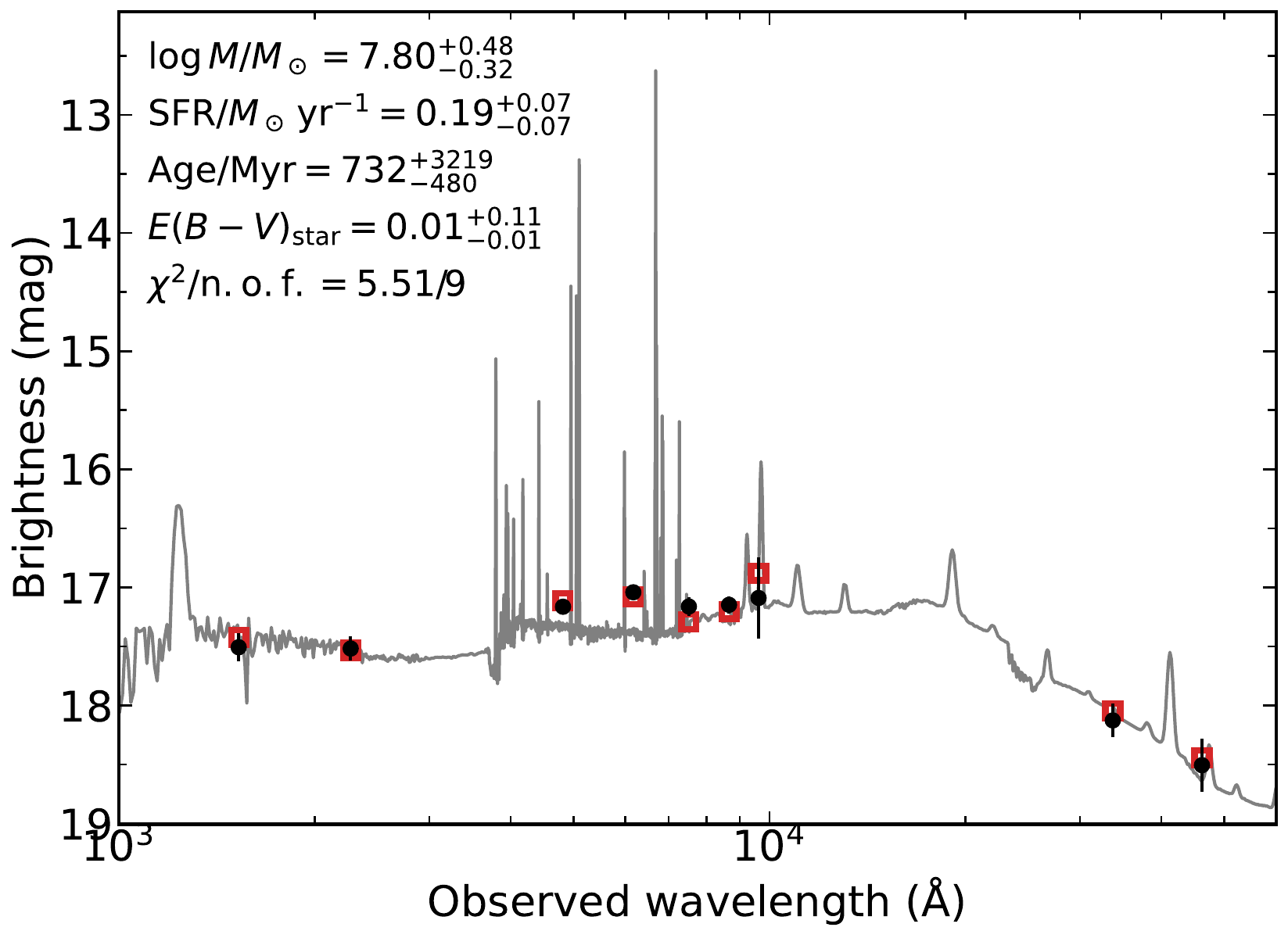}
    \caption{Flux from the host, including the flux from the bright head and the diffuse tail.}
    \label{fig:prospector_host}
  \end{subfigure}
  \hfill
  \begin{subfigure}{\columnwidth}
    \centering
    \includegraphics[width=0.8\columnwidth]{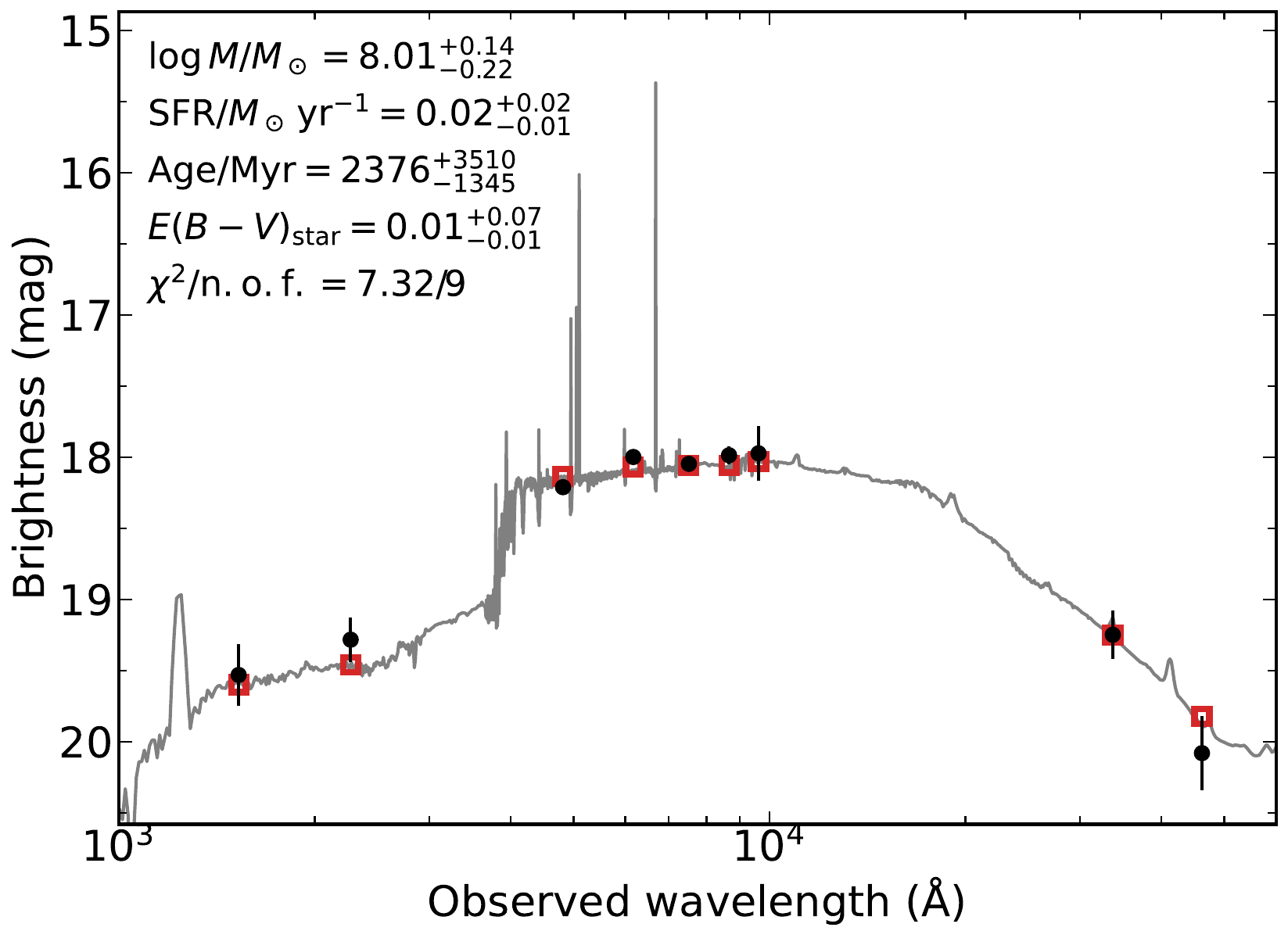}
    \caption{Flux from diffuse tail to the south-west.}
    \label{fig:prospector_head}
  \end{subfigure}
  \hfill
  \begin{subfigure}{\columnwidth}
    \centering
    \includegraphics[width=0.8\columnwidth]{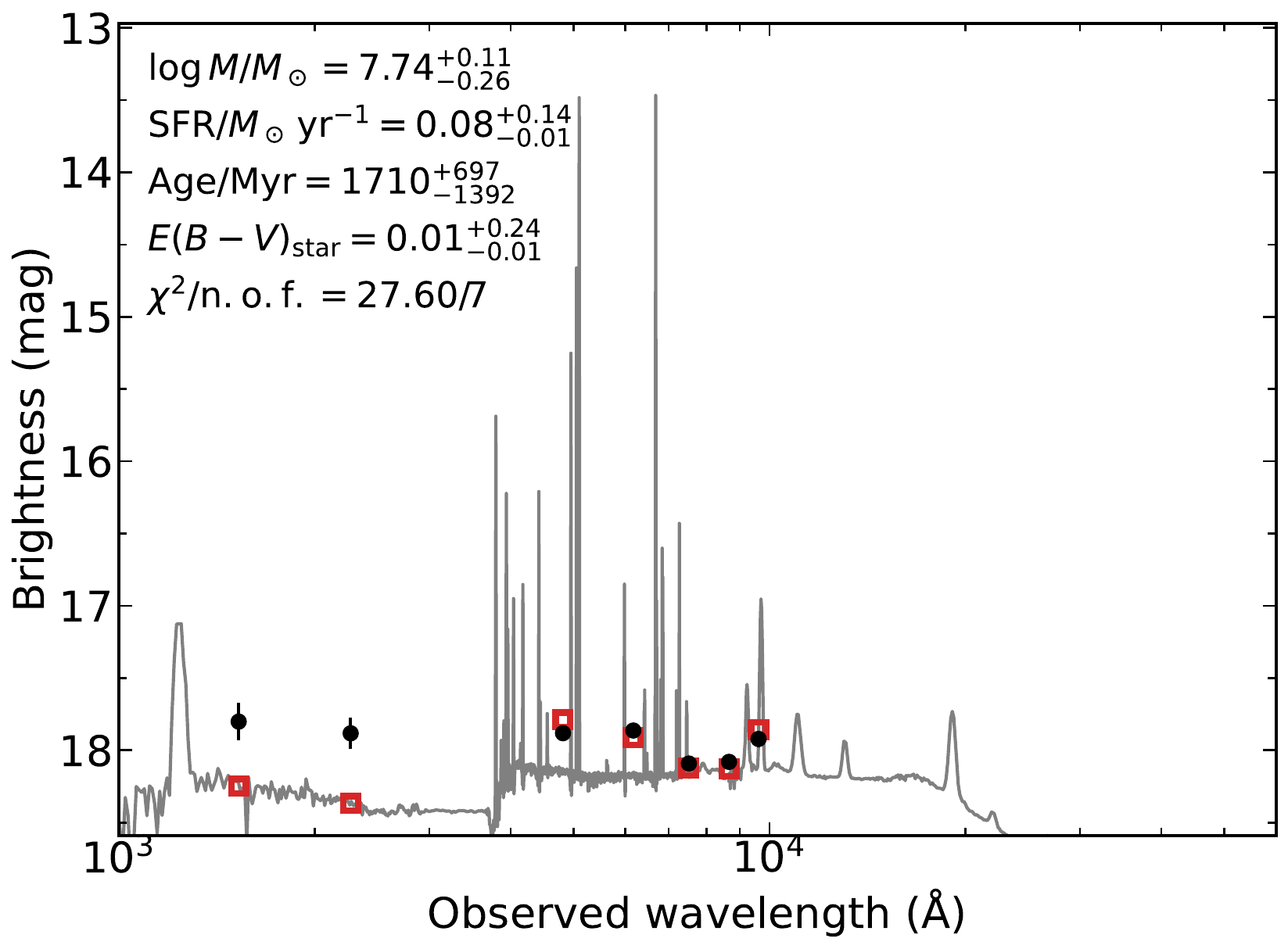}
    \caption{Flux from bright star forming head to the north-west.}
    \label{fig:prospector_tail}
  \end{subfigure}
  \caption{Spectral energy distribution of the host galaxy from 1000 to 60,000~\AA\ (black dots). The solid line displays the best-fitting model of the SED. The red squares represent the model-predicted magnitudes. The fitting parameters are shown in the upper-left corner. The abbreviation `n.o.f.' stands for the number of filters. We perform measurements on the entire host flux (upper panel), the faint tail (middle panel) and the star forming blob (lower panel) where \xdos exploded.}
  \label{fig:three_images}

\end{figure}

We model the observed spectral energy distribution (black data points in Figure \ref{fig:three_images}) with the software package \software{Prospector} version 1.1 \citep{Johnson2021a}.\footnote{\software{Prospector} uses the \software{Flexible Stellar Population Synthesis} (\software{FSPS}) code \citep{Conroy2009a} to generate the underlying physical model and \software{python-fsps} \citep{ForemanMackey2014a} to interface with \software{FSPS} in \software{python}. The \software{FSPS} code also accounts for the contribution from the diffuse gas based on the \software{Cloudy} models from \citet{Byler2017a}. We use the dynamic nested sampling package \software{dynesty} \citep{Speagle2020a} to sample the posterior probability.} We assume a Chabrier IMF \citep{Chabrier2003a} and approximate the star formation history (SFH) by a linearly increasing SFH at early times followed by an exponential decline at late times [functional form $t \times \exp\left(-t/\tau\right)$, where $t$ is the age of the SFH episode and $\tau$ is the $e$-folding timescale]. The model is attenuated with the \citealt{Calzetti2000a} model. The priors of the model parameters are set identically to those used by \citealt{Schulze2021a}. Figure \ref{fig:three_images} shows the observed SED (black data points) and its best fit (grey curve). The SED is adequately described by a galaxy template with a log mass of $7.80^{+0.48}_{-0.32}~M_\odot$, a star-formation rate of $0.19^{+0.07}_{-0.07}~M_\odot\,{\rm yr}^{-1}$. The mass and the star-formation rate are comparable to common star-forming galaxies of that stellar mass \citep[grey band in Figure \ref{fig:SFR-mass};][]{Elbaz2007a}, and they are also similar to those of the host galaxy populations of SNe IIn and SLSNe-IIn from the PTF survey \citep{Schulze2021a} albeit in the lower half of the mass distribution. In the same figure, we also show the location of the star-forming knot where the SN exploded in a lighter shade. This region has a somewhat lower specific star-formation rate (i.e., star-formation rate normalised by galaxy mass) than of the entire galaxy but consistent within errors.

\begin{figure*}
    \centering
    \includegraphics[width= \textwidth]{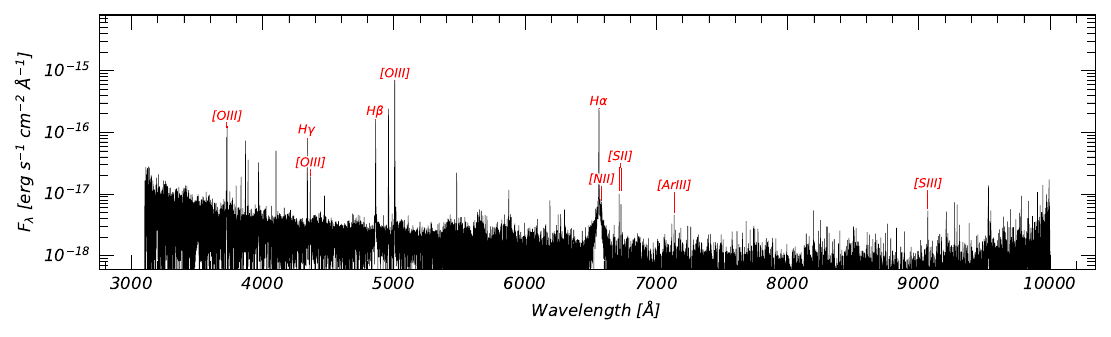}
    \caption{Spectrum extracted from a region offset from the \xshooter spectrum taken on 2023-02-26. Emission lines used in metallicity, {\rm  $T_e$} and {\rm  $n_e$} measurements are marked in red. We note transient flux is still present in the spectrum, as is seen by the broad appearance of \halpha and \hbeta.  }
    \label{fig:spectrum_host}
\end{figure*}

We focus on the narrow emission lines and the auroral emission lines associated with the underlying host \citep[see review by ][ and references therein]{Kewley2019}. The final \xshooter spectrum is utilised due to its long exposure time, and therefore higher S/N, although many of these diagnostic lines are present in all \xshooter spectra. As seen in Fig.~\ref{fig:X-shooter_latestspectra}, the late time \xshooter spectra contain numerous narrow emission lines that are typically associated with \ion{H}{II} regions. These narrow lines can provide insight into the metallicity (Z), the amount of dust, the electron temperature (${ \rm T_e } $) and density ($ {\rm n_e }$), the age of the nebula, and the rate of star formation. However transient flux is still clearly present in our latest \xshooter spectrum. We extract a trace in the 2D spectrum from a region offset from \xdos in order to minimise the contamination from \xdos, while still  detected weak lines needed for abundance measurements, given in Fig. \ref{fig:spectrum_host}. A single narrow Gaussian emission line is fitted to each emission feature, while simultaneously fitting a pseudo-continuum. This offset spectrum still displays a weak offset broad component seen in \halpha, so the following measurements (i.e those that rely on hydrogen emission lines) may be slightly overestimated due to transient contamination. Investigating the Balmer decrement, we find \halpha/\hbeta $\approx 2.5\pm0.2$ and \hgamma/\hbeta $\approx 0.47\pm0.07$, consistent with negligible interstellar reddening, as would be expected form the blue appearance of the surrounding environment, see Fig.~\ref{fig:finder_host}. 

\begin{figure}
    \centering
    \includegraphics[width= \columnwidth]{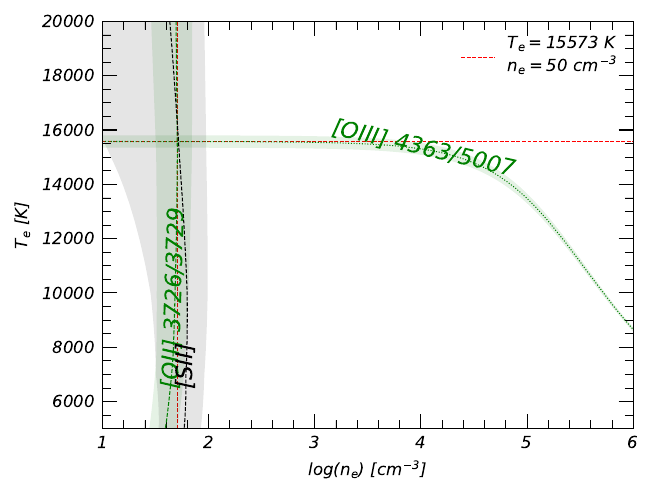}
    \caption{Emission-line temperature and density diagnostic plot from \software{PyNeb} using [\ion{O}{III}] $\lambda4363/\lambda5007$ as a temperature sensitive probe, and [\ion{O}{II}] $\lambda3726/\lambda3729$, and [\ion{S}{II}] $\lambda6731/\lambda6716$ as density-sensitive probes.}
    \label{fig:PYNEB}
\end{figure}

We measure values for {\rm  $T_e$} and {\rm  $n_e$} using \software{PyNeb}\footnote{\href{http://research.iac.es/proyecto/PyNeb/}{http://research.iac.es/proyecto/PyNeb/}} \citep{Luridiana2014} using [\ion{O}{III}] $\lambda4363/\lambda5007$ as a temperature sensitive probe, and [\ion{O}{II}] $\lambda3726/\lambda3729$, and [\ion{S}{II}] $\lambda6731/\lambda6716$ as density-sensitive probes, see Fig.~\ref{fig:PYNEB}. Encouragingly, we find similar results as for the abundance indicators above with {\rm $T_e\approx 16000~K$}, {\rm $n_e\approx 55~$cm$^{-3}$}. Using [\ion{O}{II}] and [\ion{O}{III}], we can constrain {\rm  $T_e$} and {\rm  $n_e$} well, whereas the {\rm  $n_e$} is not well constrained by [\ion{S}{II}], again likely due to the low density environment limit of 1.5 \citep{Osterbrock2006}. We search for other features to use within \software{PyNeb}, but found no other significant diagnostic lines. These may be due to the high ionising host flux, inhibiting these emission lines, or due to the low inferred densities. Using \software{PyNeb}, we find a metallicity of 12 + log((O$+$ + O$++$/H) = 7.61~dex (0.08~$Z_{\odot}$), taking the solar value to be $ { \rm 12+\log({\rm O/H})_{\odot} = 8.69 }$~dex \citep{Asplund2021} and assuming the \ion{O}{ii} temperature dependence from \citealt{Stasinska1982}.

To compare the explosion site with those of other SNe IIn, we also investigate abundance indicators using the narrow \halpha and \hbeta components, such as the commonly adopted N2 and O3N2 techniques \citep{Pettini2004}. These methods use the [N II]$~\lambda~6583$ emission line which is detected in our +480d \xshooter spectrum, but has a low $S/N\lesssim5$, and has complex underlying flux, due to the broad \halpha profile.  Although there is some scatter in the derived metallicities, all methods strongly point towards a very low metallicity environment of $0.05-0.2~Z_{\odot}$.

Tables~\ref{tab:lineflux} and \ref{tab:metallicty} give the results from the various abundance methods to determine the metallicity of the underlying environment around \xdos. Similar results (i.e. low metallicity) are found using the transient dominated spectrum from 2023-02-26 (e.g. Fig. \ref{fig:X-shooter_latestspectra}), although to irrefutably  confirm the metallicity of the host, we require late time spectral templates once \xdos has faded.

\begin{table*}
\centering
\caption{Measured parameters for several emission lines in the VLT/XShooter spectrum taken on 2023-02-26 (+480d). Flux has been corrected for MW extinction. Errors for each measurement are given in parentheses. Errors on the Equivalent Width (EW) are 
from %
\citet[][their Eq. 6]{Cayrel1988}. Note for the Hydrogen emission lines, we measure the flux from the narrow component only.}
\label{tab:lineflux}
\begin{tabular}{l|ccr}
\toprule
 & Flux [$ { \rm erg~s^{-1}~cm^{-2} }$] & FWHM [$ { \rm km~s^{-1}] }$  & EW [$ { \rm \AA] }$  \\
\midrule
$ {\rm H~\lambda4340 }$ & 3.834 (0.020) $\times 10^{-16}$ & 36.408 (0.164) & -12.790 (0.003) \\
$ {\rm H~\lambda4861 }$ & 8.351 (0.019) $\times 10^{-16}$ & 37.061 (0.156) & -19.417 (0.002) \\
$ {\rm H~\lambda6563 }$ & 2.047 (0.005) $\times 10^{-15}$ & 27.633 (0.182) & -14.439 (0.001) \\
$ {\rm [Ar~III]~\lambda7135 }$ & 2.246 (0.093) $\times 10^{-17}$ & 20.292 (0.332) & -2.494 (0.028) \\
$ {\rm [N~II]~\lambda6583 }$ & 1.495 (0.245) $\times 10^{-17}$ & 19.677 (0.626) & -0.159 (0.109) \\
$ {\rm [O~III]~\lambda4363 }$ & 6.503 (0.199) $\times 10^{-17}$ & 34.855 (0.327) & -2.242 (0.020) \\
$ {\rm [O~III]~\lambda5007 }$ & 3.079 (0.005) $\times 10^{-15}$ & 34.975 (0.202) & -139.444 (0.001) \\
$ {\rm [O~II]~\lambda3726 }$ & 4.396 (0.043) $\times 10^{-16}$ & 37.360 (0.158) & -13.415 (0.006) \\
$ {\rm [O~II]~\lambda3729 }$ & 6.253 (0.087) $\times 10^{-16}$ & 35.613 (0.209) & -19.163 (0.009) \\
$ {\rm [S~ III]~\lambda9069 }$ & 5.686 (0.095) $\times 10^{-17}$ & 22.275 (0.157) & -8.518 (0.011) \\
$ {\rm [S~ II]~\lambda6716 }$ & 7.301 (0.119) $\times 10^{-17}$ & 24.933 (0.558) & -6.277 (0.011) \\
$ {\rm [S~II]~\lambda6731 }$ & 5.263 (0.102) $\times 10^{-17}$ & 24.194 (0.472) & -4.766 (0.013) \\
\bottomrule
\end{tabular}
\end{table*} 
\begin{table}
\centering
\caption{Metallicity measurements for \xdos using verious abundance indictures found in literature.}
\label{tab:metallicty}
\begin{tabular}{lccc}
\toprule
 & $ { \rm 12 + log(O/H) } $ & $ {\rm Z/Z_{\odot}}$ \\
\midrule
PyNeb & 7.61 & 0.08 \\
N2 & 7.76 & 0.12 \\
O3N2 & 7.95 & 0.18 \\
\bottomrule
\end{tabular}
\end{table} 
\begin{figure}
    \centering
    \includegraphics[width= \columnwidth]{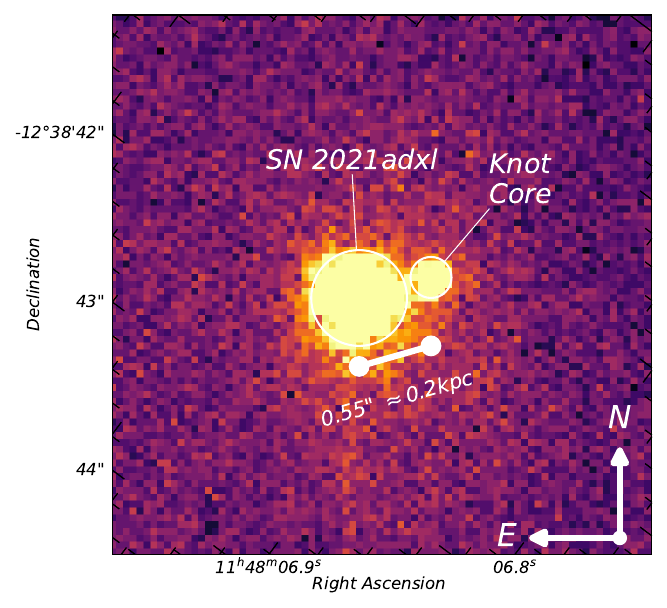}
    \caption{A 4" $\times$ 4" cutout from the HST/STIS acquisition images from 2022-04-11 (+159~d) including \xdos as well as the compact bright star forming knot. \xdos is separated from this compact knot by roughly half an arcsecond, or a projected distance of $\sim$0.2~kpc.}
    \label{fig:finder_COS}
\end{figure}

\section{Modelling the \texorpdfstring{\halpha}{} profile}\label{sec:halpha_modelling}

In order to understand the formation of the hydrogen emission lines, we note two distinct features, one is the offset flat topped emission profile, which suggests of a thin emitting shell \citep{Jerkstrand2017} and the second is the red-ward, extended wings, which are characteristic of electron scattering \citep{Huang2018}.

The collision of the  fast moving ejecta with the slower moving CSM results in the formation of a dense shell of material. This shell is made up of swept-up, shock-heated gas from the circumstellar medium and can efficiently cool to relatively low temperatures compared to the high-energy, shock-heated gas found in other parts of the supernova ejecta. This cool dense shell (CDS) of material is bound between the outward moving forward shock and receding reverse shock \citep{Chugai2004,Smith2017}. Diffusion of radiation from this shocked shell produces the main continuum (e.g. red line of Fig.~\ref{fig:earliest_spectrum}) as well as the intermediate ($\sim10^3$\kms) components of \halpha. 

Under the assumption that the \halpha profile arises from photons emitted from the CDS that undergo multiple scattering events as they travel outwards through the surrounding media, a model of electron scattering is explored. A Monte Carlo code was developed\footnote{\href{https://github.com/Astro-Sean/escatter}{https://github.com/Astro-Sean/escatter}} in order to follow photons emitted from a thin shell as they travel through a diffuse medium. The code is based on the work by \citealt{Pozdnyakov1983} and similar codes have been employed in previous works on interacting SNe \citep[e.g.,][]{Fransson2014,Taddia2020}. In this model,  photons are emitted with an emissivity $\eta$, which varies with the electron density as $\propto n_e^2$ and $n_e \propto r^{-2}$. Photons are followed until they escape the scattering medium, with a probability $\propto e^{-\tau_e}$.  We include an occulting photosphere, meaning that any escaping photons moving away from the observer are excluded. The optical depth, electron temperature, and radius are degenerate in these electron scattering models \citep[see ][for details]{Huang2018}, and are left as restricted parameters. While this is nonphysical for a SN explosion as these parameters evolve with time, the models can be used to understand the formation of the \halpha profile, as well as to further understand the main properties of the SN and, in particular, the shock velocity ($V{ \rm _{shock} } $) and mass-loss rate (\mdot).

\begin{figure}
    \centering
    \includegraphics[width= \columnwidth]{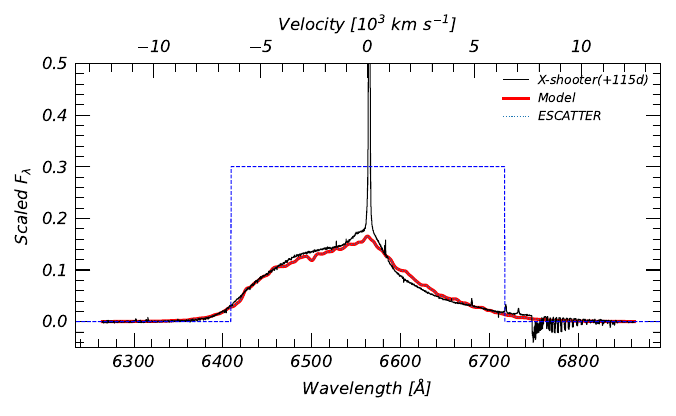}
    \caption{\halpha profile from the VLT/\xshooter spectrum (black) with the best fitting electron scattering model (red) for a given input spectrum (blue). The model matches the blue and red sides of the emission line well, but leaves an intermediate and narrow emission component at the  central wavelength. }
    \label{fig:halpha_scatter_singlefit}
 \end{figure}

Figure~\ref{fig:halpha_scatter_singlefit} shows the first epoch \xshooter spectrum with the best fitting electron scattering model. The model matches the spectrum well, capturing the blue shoulder as well as the extended red wing. However, the model fails to capture the features around rest wavelength. This likely represents emission from slowly moving material surrounding the SN ejecta, as well as emission from the host.

\begin{figure}
    \centering
    \includegraphics[width= \columnwidth]{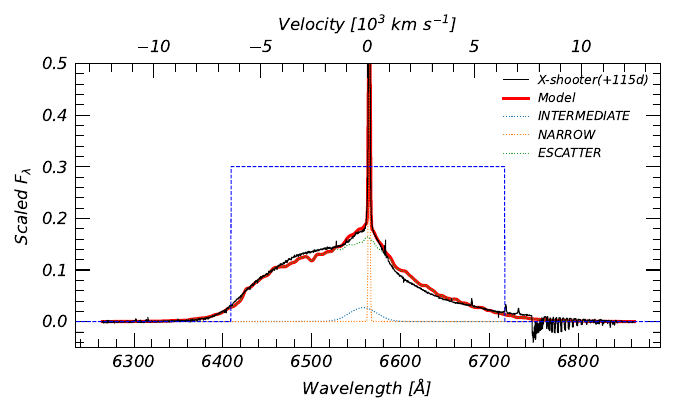}
    \caption{Same as Fig.~\ref{fig:halpha_scatter_singlefit}, but now including an additional intermediate and narrow Gaussian emission component. These three components can reproduce the entire \halpha profile well.}
    \label{fig:halpha_scatter_multifit}
\end{figure}

Similar to Fig. \ref{fig:halpha_scatter_singlefit}, Fig.~\ref{fig:halpha_scatter_multifit} shows the best fitting model for the earliest \xshooter spectrum, but with additional two Gaussian components, which likely reflect emission from the host and from CSM material far away from the shock (which we do not include in our models). As demonstrated in Fig.~\ref{fig:halpha_evolution}, the \halpha profile evolves with time, and most notably, its blueward component moves toward the central wavelength. Using a grid of electron scattering  models with an input spectrum representing a thin emitting shell, we are able to match the best fit model to the \halpha profile for each spectrum (excluding the SEDM spectra due to their low resolution) to better understand how this blueward feature evolves. Figure~\ref{fig:shock_velocity} gives the results of the evolution of the velocity needed to produce the \halpha profile at each epoch. During Phase 1, the velocity evolution is well fitted by:

\begin{figure}
    \centering
    \includegraphics[width= \columnwidth]{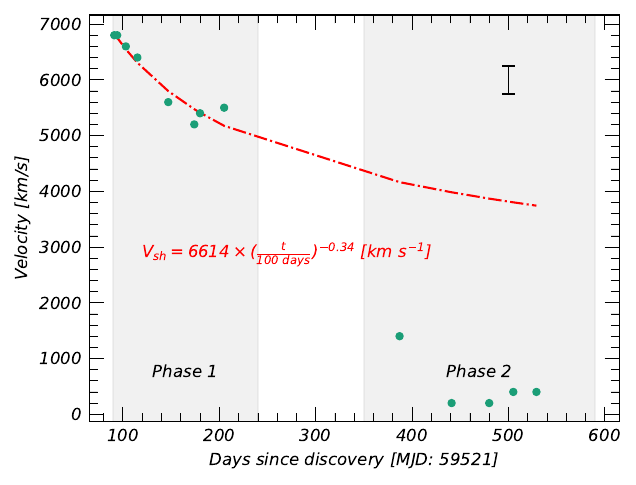}
    \caption{Input velocities of the best-fitting models needed to fit the \halpha profile evolution. Velocities during Phase 1 are fitted with an exponential decline, given in red. During Phase 2, the velocities are lower and no longer follow the trend seen in Phase 1. This is likely due to the forward shock becoming optically thin, which is also seen in the faster decay rate in the bolometric lightcurve seen in Fig. \ref{fig:lightcurve_comparision}. The errorbar in the upper right denotes the velocity grid-resolution of the input spectra. }
    \label{fig:shock_velocity}
\end{figure}

\begin{equation}\label{eq:shock_velocity}
    \rm{V_{shock}} = 6625 \times \left[ \frac{t}{100~\rm{days}} \right] ^{-0.37} ~\rm{km~s^{-1}}.
\end{equation}

\noindent At the beginning of Phase 2, the model velocities are much lower and no longer follow Eq. \ref{eq:shock_velocity}. Models fitted in Phase 2 show more scatter in their inferred velocity, and one should note the additional complexity of the \halpha profile at this time as shown in Fig. \ref{fig:halpha_evolution}. 

\subsection{Modelling the ejecta-CSM interaction}\label{ssec:modelling}

The dominating energy input during the evolution of Type IIn SNe is due to the interaction between the SN ejecta and the surrounding CSM \citep{Dessart2015}. The diversity of the mass, velocities, and geometry of this interacting material are likely responsible for the heterogeneous appearance of this subclass \citep{Chatzopoulos2012,Nyholm2020,Khatami2023}.  The total luminosity generated by ejecta-CSM interaction can be high because a radiative shock is a very efficient engine to convert kinetic energy into radiation at optical wavelengths. The luminosity of the CSM interaction is dependent on the velocity at which CSM enters the forward shock and on the progenitor's mass-loss rate (\mdot), and is typically given by:

\begin{equation}\label{eq:shock_luminousity}
\rm{L_{shock} = 2\pi \rho_{CSM} r_{shock}^2 V_{shock}^3 = \epsilon \frac{1}{2}\frac{ \dot M }{v_{wind}} V_{shock}^3~{\rm erg~s^{-1}}}
\end{equation}

\noindent assuming a steady mass-loss rate for the CSM \citep{Wood-Vasey2004} and a 50\% conversion efficiency ($\epsilon=0.5$). It is often difficult to determine the shock velocity from the line profiles given the uncertainty to which extent electron scattering plays a role in determining the widths of the intermediate lines \citep[see discussion by][]{Smith2017}. \xdos offers a unique test to measure the degree to which electron scattering plays a role, as discussed in Sect.~\ref{sec:halpha_modelling}, and therefore to gain insight into the ejecta-CSM interaction.

\begin{figure*}
    \centering
    \includegraphics[width= \textwidth]{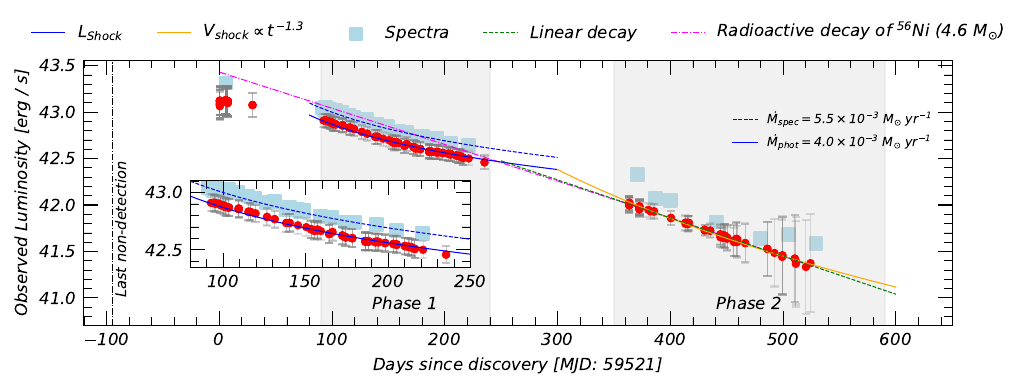}
    \caption{Pseudo-bolometric lightcurve of \xdos . Best fitting model from Eq. \ref{eq:shock_luminousity} fitting to Phase 1 photometry (spectra) is given as the blue solid (dashed) blue line. Included in the inset is a zoom in of this phase.  Both a  linear decline (dashed green) and possible shock luminosity curve (solid orange) is fitted to Phase 2.  The magenta line is the decay expected for radioactive nickel, fitted to Phase 2, as discussed in Sect. \ref{ssec:modelling}.}
    \label{fig:bolometric}
\end{figure*}

Using Eqs.~\ref{eq:shock_velocity} and \ref{eq:shock_luminousity}, we fit the Phase 1 portion of the lightcurve with the mass-loss rate  \mdot as a free parameter. The CSM velocity is taken to be $\sim$250\kms, as seen in the earliest spectrum, see Fig.~\ref{fig:FIRE_spectrum}, as well as the HST/COS spectrum given, see Fig. \ref{fig:HST_COS}. As discussed in Sect.~\ref{sec:halpha_modelling}, the blue component of the \halpha profile is likely caused by line scattering from a somewhat optically thin CDS. One interpretation of this is that photons are emitted from the outer regions of the CDS, and while travelling radially outwards, scatter off material swept up by the forward shock. 

It is important to note that the measurements of the shock velocity as provided in Eq.~\ref{eq:shock_velocity} are not based on the photometric evolution, but rather model fitting of the \halpha profile. The exponent for the shock velocity evolution in Eq. \ref{eq:shock_luminousity} is responsible for the slope, and it is encouraging that the resulting fit in Fig.~\ref{fig:bolometric} matches the bolometric evolution so well. Equations \ref{eq:shock_velocity} and \ref{eq:shock_luminousity} fits the luminosity of Phase 1  but this does not extend to Phase 2, where a faster, apparently linear, decline is seen.  Extrapolating both trends, we expect a break around +300~d. Following Eq. \ref{eq:shock_luminousity}, a steady state mass-loss rate of $4-6\times10^{-3}$~\msunperyr is required to provide the necessary luminosity if powered solely by shock interaction. Additionally, it is possible to fit a second decline similar to Eq. \ref{eq:shock_luminousity} to Phase 2 with a quicker velocity decline with an exponent of $\approx-1.3$, however it is difficult to constrain this.

The use of Eq. \ref{eq:shock_luminousity} is likely an oversimplification but provides an estimate to the the mass-loss rate from the progenitor shortly before the SN explosion. A mass-loss rate of $10^{-3}$~\msunperyr is consistent with reported values for other interacting SNe \citep{Taddia2013,Ofek2014,Fransson2014,Moriya2014,Moriya2023} as well as the values often assumed for Luminous Blue Variable (LBV) progenitors \citep{Trundle2008, Dwarkadas2011,Groh2013, Smith2017b,Weis2020} as discussed further in Sect.~\ref{ssec:progenitor}.

As shown in Fig. \ref{fig:multiband_lightcurve}, \xdos undergoes a steeper decline after Phase 1 and the trend seen in Fig. \ref{fig:shock_velocity} may be related, perhaps reflecting a change in CSM density, distribution, or opacities,e.g. see discussion by for SN~2010jl \citep{Ofek2014,Moriya2014b}. This is also observed in the evolution of the \halpha profile in Fig.\ref{fig:halpha_evolution}, when the overall profile narrows during Phase 2, likely reflecting a lower optical depth for material ahead of the shock from, meaning less scattering events \citep{Huang2018} and less broading of the emission profile.

If we assume the ejecta-CSM interaction ceases when the shock reaches the edge of the dense CSM at a time ${\rm t_{end}}$, we can estimate the radial extent of this material (${\rm d_{CSM}}$) using:

\begin{equation}\label{eq:distance_intgral}
{ \rm d_{CSM} = \int_{t_{start}}^{t_{end}}  (t - t_{start}) \times V_{shock} (t) \,dt},
\end{equation}

\noindent where ${ \rm t_{start}}$ is the time at which the SN ejecta collides with the CSM, launching the forward shock into the dense CSM, and we use  ${\rm V_{shock}(t)}$ from Eq. \ref{eq:shock_velocity}. We do not have a constraint on the explosion date so we can not be certain on the duration of CSM interaction. 

We make an conservative estimation of ${\rm d_{CSM}}$ by assuming that ${\rm t_{start}=100}$~days, ${\rm t_{end}=300}$~days and ${\rm t_{duration} \approx 200}$~days (although this value may be a factor of 2 or more greater), the timeframe for which Eq. \ref{eq:shock_velocity} is likely to be valid i.e the duration of Phase 1. This gives an approximate size of ${ \rm d_{CSM}}$ of $8\times10^{17}$~cm (this is more than an order of magnitude larger that that reported for SN~2010jl \citealp[ for example ][]{Dwek2021}) meaning that the CSM travelling at $\sim250~$\kms would have been ejected in the last $\sim$1000 years. We can set a lower limit on the CSM mass in Phase 1, assuming a constant \mdot$=4\times10^{-3}$~\msunperyr, homologous distribution, and spherical symmetry, as ${\rm M_{CSM}\gtrapprox~4.3}$~\msun. Additionally this mass will be substantially higher if we have overestimated ${\rm V_{wind} }$ or underestimated ${\rm t_{start}}$ and ${\rm t_{end}}$, although this is hard to quantify as we are uncertain how ${\rm V_{shock}}$ evolves for $t\lesssim100$~days. This high CSM mass is similar to that measured for SN~2010jl \citep[$\sim1-10$~\msun;  ][]{Zhang2012,Fransson2014,Ofek2014} and is likely a contributing reason for the slow evolution seen in both transients.

Assuming \xdos to be entirely powered by radioactive decay, Fig. \ref{fig:bolometric} includes the energy deposited by a mass of explosively synthesised radioactive material given by \citealt{Jeffery1999}, and the mass of radioactive nickel needed would be on the order of 3-10~\msun of material. However such a mass of nickel would produce a much brighter transient, and would over-predict the luminosity evolution during Phase 1. This is obviously an overestimate, and not expected for \xdos, as the transient is likely dominated by CSM interaction even in Phase 2. We conclude that CSM interaction likely dominates Phase 2, although there is a different CSM mass / distribution when compared to Phase 1 meaning the progenitor has undergone several mass loss episodes in the final moments before core-collapse.

\section{Discussion}\label{sec:discussion}

\begin{figure*}
    \centering
    \includegraphics[width= \textwidth]{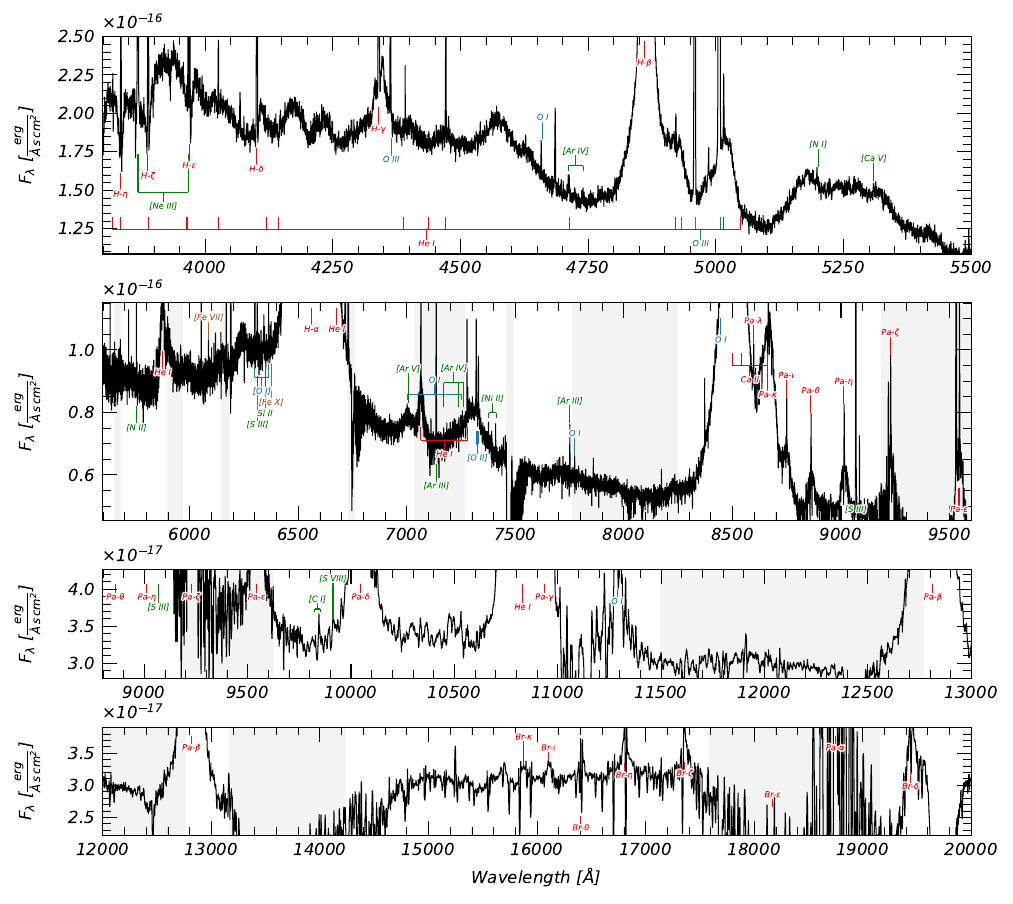}
    \caption{ VLT/\xshooter for \xdos taken on 2023-02-26  (+480d). The spectrum has been fluxed calibrated to the $r$-band and each panel highlights a wavelength range which (roughly) covers the near UV, optical, and near-IR. Telluric bands are marked by the grey area. The bottom two panels have been smoothed using a Savgol filter for clarity.}
    \label{fig:X-shooter_latestspectra}
\end{figure*}

\xdos is an example of a nearby, bright, long-lasting interacting supernova. This allows for a rare glimpse to the late-time evolution of the transient and for this purpose we have obtained high-resolution spectra. Figure~\ref{fig:X-shooter_latestspectra} provides a detailed picture of the final VLT/\xshooter spectrum taken at +480d. This highlights the weaker, narrow emission features that are typically missed when lower-resolution spectrographs are employed. For context, many narrow features discussed in this section are not detected in the NOT+ALFOSC spectra. This suggest an inefficient observing strategy for late time transient followup when only low-medium spectrum are obtained.

\begin{figure}
    \centering
    \includegraphics[width= \columnwidth]{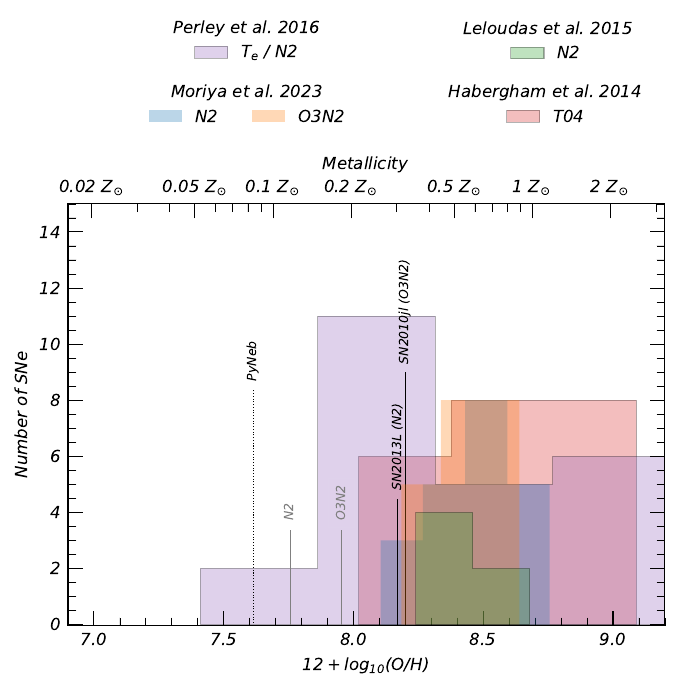}
    \caption{Metallicity measurements for the environment around \xdos using the VLT/\xshooter spectrum on 2023-02-26 (+480d). Histograms are measured metallicities for hosts of Type IIn SNe \citep{Habergham2014,Moriya2023}, Type II SLSNe \citep{Leloudas2015a}, as well as both Type I and II SLSNe \citep{Perley2016} using various methods, denoted in the legend, and detailed in Sect.~\ref{ssec:host}. Vertical lines mark measured values for \xdos using different methods (see text). We also mark the reported metallicity for SN~2010jl at $Z \approx 0.32~Z_{\odot}$ \citep{Stoll2011} and for SN~2013L at $Z \approx 0.30~Z_{\odot}$ \citep{Taddia2020}. }
    \label{fig:metallicity_histogram}
\end{figure}

\subsection{The nature of the \xdos's host }\label{ssec:green_pea}

The \lalpha profile seen in Fig. \ref{fig:HST_COS} is in contrast to the appearance of \halpha as shown in Fig. \ref{fig:X-shooter_halpha_evolution}. While this may be an opacity effect, or possibly due to emission from material far away from the ejecta-CSM interface as mentioned above, similar profiles for \lalpha have been seen for population of compact,  star-forming galaxies at z $\sim$ 0.2 - 0.3, also known as ``green peas'' (GP) galaxies \citep{Cardamone2009}. These GP galaxies have a high  $ {\rm [\ion{O}{III}]~\lambda~5007 }$  / $ {\rm [\ion{O}{II}]~\lambda~3726 }$  ratio, high equivalent width in \ion{O}{III} emission lines \citep[See Table. \ref{tab:lineflux}][]{Cardamone2009}, and are typically sub- solar metallicity  \citep{Orlitov2018,Liu2022}. Our \software{Prospector} measurements also find a mass consistent with GP \citep[$\sim10^8$~\msun;][]{Cardamone2009}. Although due to its proximity, the host of \xdos would be classified as a Blueberry galaxy rather than a Green Pea \citep{Liu2022}.

As shown in Fig. \ref{fig:HST_COS}, the \lalpha shows a unique profile that is not clearly seen in any other emission profile. The profile may resemble a P-Cygni emission profile, typically observed in stellar winds and SNe, or a doubled peaked emission profile with an absorption trough at center wavelength. We note the similarities of the HST/STIS \lalpha profile to that observed in GP galaxies \citep[e.g.][]{Yang_2016,Orlitov2018,Henry2018}. In this case, the asymmetric profile is thought to be due to resonant scattering of \lalpha photons off neutral hydrogen atoms, which is dependant on the intervening material and fraction of escaping ionizing radiation.  This would also point that the broad feature in Fig. \ref{fig:HST_STIS} would be \ion{Mg}{II} $\lambda\lambda$2796, 2803 \citep{Henry2018}, originating from the host and not \xdos, although does not explain the observed wavelength offset.

\subsection{Type IIn SNe exploding in metal poor environments}\label{ssec:metal_poor_host}

Perhaps greater insight may be gained from the local environment around \xdos. As mentioned in Sect.~\ref{ssec:host}, \xdos occurred in a bright star-forming region at the end of an elongated intensity distribution. We find that the local environment has a very low metallicity of ${ \rm 0.06~Z_{\odot} }$. This is similar to the environments for SN~2010jl \citep{Stoll2011} and SN~2013L \citep{Taddia2020}, as well as for other Type IIn SNe \citep{Habergham2014,Moriya2023}. 

 Figure~\ref{fig:metallicity_histogram} compares the measured metallicities for the host of \xdos to a sample of  values for hosts of Type IIn SNe \citep{Habergham2014,Moriya2023}, Type II SLSN \citep{Leloudas2015a}, as well as both Type I and II SLSNe \citep{Perley2016}. \xdos is at drastically lower metallicity than the sample of Type IIn SNe \citep[e.g. samples from ][]{Habergham2014,Moriya2023}, and also lower that for Type II SN hosts \citep[except perhaps for SN~2015bs; ][]{Anderson2016,Anderson2018}. The sample of \citealt{Moriya2023} does include SN~2016iaf having 12 + log(O/H) = $7.48\pm0.81$ using the D16 method \citep{Dopita2016}, although this metallicity is outside the reported validity range for D16.

\citealt{Moriya2023} find that Type IIn SNe with a higher peak luminosity  occur in environments with lower metallicity and young stellar environment. Although \xdos is beyond the range explored by \citealt{Moriya2023}, the location and metallicity of the environment \xdos occurred in may explain its brightness.  The conjecture that more luminous IIn's are found in less metal-enrichment environments is in conflict with results from \citealt{Leloudas2015a}, \citealt{Perley2016} and \citealt{Schulze2018,Schulze2021a}. Figure \ref{fig:metallicity_histogram} is contradicting this scenario, too. The SLSN-II sample spans the full range of metallicities \citep[e.g. sample from ][ in Fig. \ref{fig:metallicity_histogram}]{Leloudas2015a}. The fact that they are super-luminous, and therefore more luminous than \xdos implies that their environments should on average be less metal enriched than of \xdos, in contradiction with the result of \citealt{Moriya2023}. Additionally,\citealt{Moriya2023} find a strong inverse correlation between peak luminosity and metallicity, which may possibly explain the brightness of \xdos although the peak magnitude would be overestimated by almost 2 magnitudes (see Fig. 1 of \citealt{Moriya2023}).

\subsection{Progenitor of \xdos}\label{ssec:progenitor}

A major problem in supernova science concerns identifying the progenitors of Type IIn SNe. Progress is further hindered by the lack of clear nebular emission line signatures \citep[which may reveal clues to the progenitor e.g. ][]{Fraser2013,Jerkstrand2020}. We are therefore often left with modelling the lightcurve and spectra, which in the case of interacting transients, are dominated by the effects of interaction, obscuring the signatures of any explosion mechanism \citep[][]{Arcavi2017,Woosley2018}.

There are no progenitor detections of \xdos, so we must rely on indirect methods. In Sect.~\ref{ssec:modelling}, we concluded that \xdos can be powered by the ejecta colliding with a previous wind with $\dot{M}\approx10^{-3}$\msunperyr. This mass-loss rate regime is often reserved for LBVs during their outburst stages \citep{Lamers1989,Weis2001,Vink2008a,Smith2013c}, and additionally the inferred wind velocity is also similar to that observed among Galactic LBVs \citep{Weis2001}.

Such high mass-loss rates are commonly inferred for Type IIn SNe \citep{Moriya2014}, although it is difficult to have a physical mechanism that can lead to this amount of mass ejected \citep{Fuller2017,Tsang2022}. Such a high steady state mass-loss rate is unlikely to be obtained from a single star. Using Eq. 20 from \citealt{Bjorklund2021}, a massive hot star would require a luminosity of $L\approx10^{7.2}~L_{\odot}$ to achieve a steady state mass-loss rate of $10^{-3}$\msunperyr at a metallicity of 0.2~$Z_{\odot}$ (this is the lower validity range from \citealt{Bjorklund2021}), which would likely exceed the Eddington luminosity of a single star. This is similar to the luminosity expected from the 1840 Giant Eruption of Eta Carinae \citep{Davidson2012}, where it is thought that binary interaction was the cause of the eruption \citep[see ][ for a possible scenario]{Smith2011b}.

The involvement of a companion in the progenitor history is an often explored formation channel for Type IIn SNe \citep{Chatzopoulos2012, Justham2014, Zapartas2019}, as most massive stars are expected to be born and interact with a binary companion \citep{Sana2012,Zapartas2019}. The environments of Type IIn do not show dramatic differences when compared to normal Type II\citep{Kelly2012}, suggesting there is no significant environment dependence for these two sub types, and (possibly) they are influenced by the present (or absence) of a companion. At the time of writing \xdos no obvious signatures of binarity (e.g. pre-explosion variability, multi-peaked emission lines) are detected. While \xdos is a luminous Type IIn supernova, it does not reach the often assumed -21 mag criteria of SLSNe \citep{Gal-Yam2012}. However in the context of interacting SNe,  it is not clear whether SNe IIn fulfilling this criterion constitute a separate population of events \citep{Moriya2018,Gal-Yam2019,Nyholm2020} or whether there is a continuum of possible energetics \citep[see Fig.~1 of ][]{Moriya2023}.

\section{Conclusions}\label{sec:conclusions}

We present 1.5 years of the evolution of the luminous Type IIn \xdos, including four epochs of spectrum from the VLT/\xshooter. \xdos, similar to SN~2010jl, does not evolve quickly, and since discovery has only faded by $\sim$4 magnitudes. This likely reflects that the forward shock, created from the ejecta-CSM interaction, is moving outwards through a dense, extended CSM environment. The effects of this environment are seen in the appearance of the \halpha emission line, which displays a persistent blue shoulder emission component. We find that this feature is due to electron scattering through an increasingly diffuse medium ahead of the forward shock. Using the evolution of this feature, we find that the post-peak evolution is dominated by the ejecta-CSM interaction. Assuming spherical symmetry and homologous density distribution, this translates to a shock wave moving through dense medium of $\gtrsim$3~\msun extended out to at least $\sim10^{17}$~cm. Between 250 and 350d, the properties of the CSM interaction changes. The light curve decay increased and the broad \halpha profile narrows, likely due to a decrease in the CSM density. This may signify a change in the evolution of the progenitor within the last $\sim10^3$ years.

Perhaps one of the most striking features of \xdos is where it occurred. The vicinity around \xdos displayed a prominent blue appearance, and we show that this is a low density, low metallicity environment. The peak brightness of \xdos may be related to its metallicity, as more metal poor progenitors may produce bright Type IIn SNe \citep[e.g.][]{Moriya2014}. Although this is likely not the (direct) cause of the high mass loss rates, as Type IIn SNe occur in heterogeneous environments, with multiple formation pathways \citep{Ransome2022}. Thanks to its proximity and (expected) slow evolution, \xdos will be observable again in early-2024, allowing for continued observations at +800 days.

\section*{Acknowledgements}

S. J. Brennan and R. Lunnan acknowledge support by the European Research Council (ERC) under the European Union’s Horizon Europe research and innovation programme (grant agreement No. 10104229 - TransPIre). S. Schulze is partially supported by LBNL Subcontract NO. 7707915 and by the G.R.E.A.T. research environment, funded by Vetenskapsr{\aa}det,  the Swedish Research Council, project number 2016-06012. T. W. Chen acknowledges the Yushan Young Fellow Program by the Ministry of Education, Taiwan for the financial support. Y.-L. Kim has received funding from the Science and Technology Facilities Council [grant number ST/V000713/1]. 

Based on observations obtained with the Samuel Oschin Telescope 48-inch and the 60-inch Telescope at the Palomar Observatory as part of the Zwicky Transient Facility project. ZTF is supported by the National Science Foundation under Grant No. AST-2034437 and a collaboration including Caltech, IPAC, the Weizmann Institute of Science, the Oskar Klein Center at Stockholm University, the University of Maryland, Deutsches Elektronen-Synchrotron and Humboldt University, the TANGO Consortium of Taiwan, the University of Wisconsin at Milwaukee, Trinity College Dublin, Lawrence Livermore National Laboratories, IN2P3, University of Warwick, Ruhr University Bochum and Northwestern University. Operations are conducted by COO, IPAC, and UW. SED Machine is based upon work supported by the National Science Foundation under Grant No. 1106171 . This work was supported by the GROWTH project funded by the National Science Foundation under Grant No 1545949. The Oskar Klein Centre is funded by the Swedish Research Council. The data presented here were obtained [in part] with ALFOSC, which is provided by the instituto de Astrofisica de Andalucia (IAA) under a joint agreement with the University of Copenhagen and NOT. The ZTF forced-photometry service was funded under the Heising-Simons Foundation grant \#12540303 (PI: Graham).The Gordon and Betty Moore Foundation, through both the Data-Driven Investigator Program and a dedicated grant, provided critical funding for SkyPortal.

\section*{Data availability}

The spectroscopic data underlying this article are available in the Weizmann Interactive Supernova Data Repository  \citep[WISeREP\footnote{\href{https://wiserep.weizmann.ac.il/}{https://wiserep.weizmann.ac.il/}}][]{Yaron2012}. The photometric data underlying this article are available in the online supplementary material.

\bibliographystyle{aa}
\bibliography{references.bib} %

\end{document}